\documentclass[prl,aps,twocolumn,showpacs,preprintnumbers,amsmath,amssymb]{revtex4}

\usepackage{graphicx}
\usepackage{dcolumn}
\usepackage{color}

\begin{document}

\def\EF{$E_\textrm{F}$}
\def\cred{\color{red}}
\def\cblue{\color{blue}}
\definecolor{dkgreen}{rgb}{0.31,0.49,0.16}
\def\cgreen{\color{dkgreen}}

\title{Phase stability and large in-plane resistivity anisotropy in
the 112-type iron-based superconductor Ca$_{1-x}$La$_{x}$FeAs$_{2}$}

\author{Chang-Jong Kang}
\affiliation{
Department of Physics and Astronomy, Rutgers University,
Piscataway, New Jersey 08854, USA
}
\author{Turan Birol}
\affiliation{
Department of Physics and Astronomy, Rutgers University,
Piscataway, New Jersey 08854, USA
}
\affiliation{
Department of Chemical Engineering and Materials Science, University of Minnesota,
Minneapolis, Minnesota 55455, USA
}
\author{Gabriel Kotliar}
\affiliation{
Department of Physics and Astronomy, Rutgers University,
Piscataway, New Jersey 08854, USA
}
\affiliation{
Condensed Matter Physics and Materials Science Department, Brookhaven National Laboratory,
Upton, New York 11973, USA
}

\date{\today}

\begin{abstract}
The recently discovered high-T$_c$ superconductor Ca$_{1-x}$La$_{x}$FeAs$_{2}$ is a unique compound
not only because of its low symmetry crystal structure,
but also because of its electronic structure which hosts Dirac-like metallic bands
resulting from (spacer) zig-zag As chains.
We present a comprehensive first principles theoretical study of the electronic and crystal structures
of Ca$_{1-x}$La$_{x}$FeAs$_{2}$.
After discussing the connection between the crystal structure of the 112 family, which Ca$_{1-x}$La$_{x}$FeAs$_{2}$
is a member of, with the other known structures of Fe pnictide superconductors,
we check the thermodynamic phase stability of CaFeAs$_{2}$,
and similar hyphothetical compounds SrFeAs$_{2}$ and BaFeAs$_{2}$
which, we find, are slightly higher in energy.
We calculate the optical conductivity
of Ca$_{1-x}$La$_{x}$FeAs$_{2}$ using the DFT + DMFT method,
and predict a large in-plane resistivity anisotropy in the normal phase,
which does not originate from electronic nematicity, but is enhanced
by the electronic correlations.
In particular, we predict a 0.34 eV peak in the $yy$ component of the optical conductivity
of the 30\% La doped compound,
which correponds to coherent interband transitions
within a fast-dispersing band arising from the zig-zag As-chains
which are unique to this compound.
We also study the Landau free energy for Ca$_{1-x}$La$_{x}$FeAs$_{2}$
including the order parameter relevant for the nematic transition
and find that the free energy does not have any extra terms that could
induce ferro-orbital order.
This explains why the presence of As chains does not broaden the nematic transition
in Ca$_{1-x}$La$_{x}$FeAs$_{2}$.

\end{abstract}


\maketitle

\section{Introduction}

The discovery of high temperature superconductivity
in the iron based materials \cite{Kamihara08}
triggered a large number of investigations
\cite{Basov11,Hirschfeld11,Stewart11,Charnukha14}.
The basic ingredient of this class of materials is square nets of iron
tetrahedrally coordinated by a pnictide or a chalcogenide.
By now this ingredient has been realized in multiple prototypical structures, such as
the original realization in the 1111 structure,
the 111 structure of LiFeAs, the 11 structure such as that of FeSe,
and the 122 structure such as in BaFe$_{2}$As$_{2}$.

The possibility of superconductivity in iron based compounds with a 112 structure
was suggested based on electronic structure calculations in Ref. \cite{Shim09},
where it was pointed out that this structure
would support metallic spacer layers which could aid in elucidating the
mechanism for high temperature superconductivity.
Attempts to synthesize iron pnictide compounds in this structure were not originally
successful, but new Mn based materials in this structure were found \cite{Wang11,Park11} and it was observed theoretically \cite{Wang11} and experimentally \cite{Park11,Jo14,Jia14,Feng14} that the spacer layers exhibit Dirac cones \cite{Lee13}.
Recently, Fe superconductors in the 112 structure were synthesized,
(Ca,Pr)FeAs$_{2}$ \cite{Yakita14}, and Ca$_{1-x}$La$_{x}$FeAs$_{2}$ \cite{Katayama13}.

These materials form in a structure where the As in the CaAs layers are distorted
in zig zag chains, i.e.
the space group $P2_{1}$ or $P2_1/m$ rather than the originally assumed tetragonal structure \cite{Katayama13,Yakita14}.
Second harmonic generation experiments confirmed the space group $P2_{1}$ for La doped compounds \cite{Harter16}.
More recently, theoretical studies focusing
on the spacer layers determined that the As $p_x$ and $p_y$
orbitals are responsible for the Dirac cones, and the spin orbit coupling
can open a gap and induce topological phases on these layers,
suggesting the 112 compounds as prime candidates for proximity induced topological
superconductivity \cite{Wu14,Wu15}.

These works motivate us to revisit the early theoretical predictions in the light
of these experimental developments to address some basic
questions. (1) The original density functional theory (DFT) calculations focused on the
FeAs layers only.
A natural question is what would be the result of a full relaxation of the crystal structure?
(2) Parent compounds such as CaFeAs$_{2}$ have so far not been synthesized,
and a rare earth is needed to facilitate the synthesis \cite{sala2014}.
Raising the question of relative stability of these compounds,
what is the role of the rare earth like La in stabilizing the structure?
(3) Photoemission experiments, confirmed the theoretical prediction of
the existence of metallic spacer layers
(with Fermi pockets of As $p_{z}$ and Ca character, in addition to the
Dirac cones) \cite{Li15,Jiang16}, however, it is not clear from them what the role
of doping is, since in the 112 structure both
the CaAs and the FeAs layers can accommodate carriers. (4) It would also be useful
to establish consequences of the anisotropy introduced by the formation
of the CaAs chains, which should be visible in the optical response,
and elucidate how this anisotropy couples to the nematic
order parameter whose origin is a subject of intensive discussion.
In this paper we answer these questions and determine how this iron pnictide fits
with the other families already studied within dynamical mean-field theory (DFT+eDMFT).
We conclude that in spite of the strong anisotropy induced by the As zig zag chains,
the 112 compounds are very similar to the rest of the iron pnictide superconductors,
indicating the superconductivity resides essentially on the FeAs layers
unaffected by spacers.

In this work, we (1) present a detailed explanation of the crystal structure of CaFeAs$_2$, and elucidate the connection of it with other pnictide superconductors, (2) check the phase stability of CaFeAs$_{2}$, as well as SrFeAs$_2$ and BaFeAs$_2$ at the level of DFT by finding the optimum first principles structure via an evolutionary structure search, and then building the convex hull, (3) systematically study the electronic structure of Ca$_{1-x}$La$_{x}$FeAs$_{2}$ using state of the art DFT + embedded DMFT (eDMFT) to understand the effect of carrier doping, and (4) for the first time predict the optical conductivity of this compound at the level of DFT + eDMFT. We show that the presence of metallic As chains on the spacer layer gives rise to a strong in-plane anisotropy even at high temperatures, distinct from that driven by the nematic transition. Comparing DFT and eDMFT results, we further show that this anisotropy is enhanced by the electronic correlations.

\begin{figure}[b]
\includegraphics[width=8.5 cm]{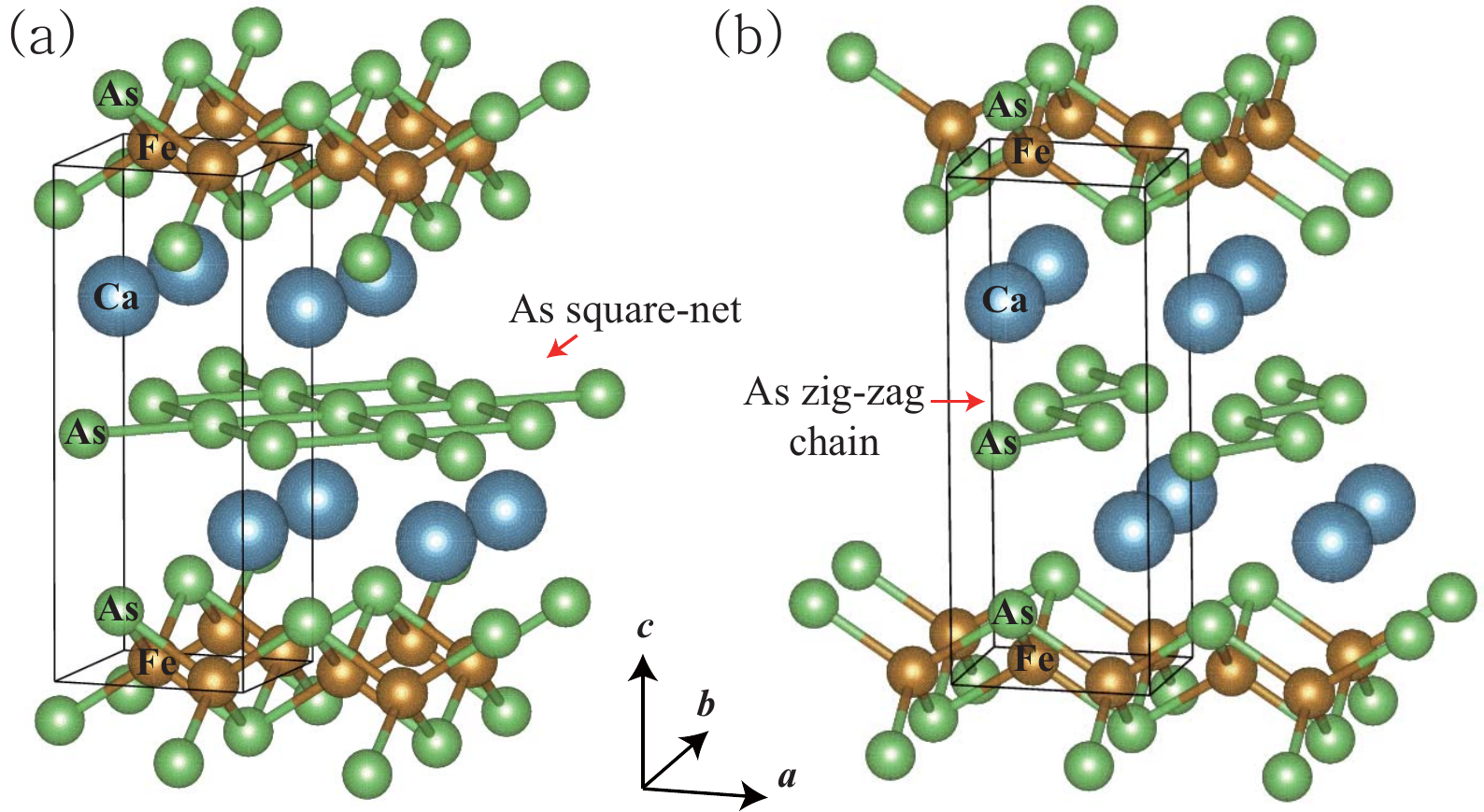}
\caption{(Color Online)
(a) Tetragonal crystal structure with $P4/nmm$ of CaFeAs$_{2}$.
(b) Monoclinic crystal structure with $P2_{1}/m$ of CaFeAs$_{2}$.
The black lines represent the unit cell in both (a) and (b).
The As square-net is clearly shown in (a), whereas,
the zig-zag As chain appears in (b).
Both spacers, As square-net and the zig-zag As chain,
are metallic.
}
\label{struct}
\end{figure}

\section{Methods}
DFT calculations are performed using the projector augmented wave method as implemented in the Vienna Ab Initio Simulation Package (VASP) \cite{VASP1, VASP2, PAW1, PAW2}. A plane wave cutoff of 500 eV and the generalized gradient approximation of Perdew-Burke-Ernzerhof (PBE) are used \cite{PBE}.

To obtain the lowest energy structures, an \emph{ab initio} evolutionary search algorithm \cite{Oganov06} as implemented in the USPEX package \cite{uspex} is used in conjunction with VASP. In these calculations, initial structures are randomly generated and the next generations are generated using mutations of the previous ones and new random structures. Once the stable structural phases are obtained from the evolutionary search, corresponding formation energies are calculated with
a dense Monkhorst-Pack sampling grid with a resolution of $2\pi \times 0.02 {\AA}^{-1}$ for the $k$-space integrations.

For the phonon dispersion calculations, the PHONOPY code \cite{phonopy} is used to build the supercells, and to find the minimum number of ionic displacements required. The \textit{direct method} is used to calculate the forces in these supercells, as opposed to the Density Functional Perturbation Theory. The force constants and the dynamical matrices are obtained from the Hellmann-Feynman forces in these (2 $\times$ 2 $\times$ 2) supercells (16 formula units).
A 6 $\times$ 6 $\times$ 2 $k$-point mesh is used in these supercell calculations.

To treat correlation effects in the Ca$_{1-x}$La$_{x}$FeAs$_{2}$ compounds,
the fully charge self-consistent scheme DFT + eDMFT \cite{Anisimov97,Savrasov04}
(for a review see Ref. \cite{Kotliar06}) is used as implemented in Ref. [\onlinecite{Haule10,eDMFT}]
with the hybridization expansion continuous-time quantum Monte Carlo \cite{Werner06}
as the impurity solver as implemented in Ref. [\onlinecite{Haule07}].
We use the Coulomb interaction $U$ = 5.0 eV and the Hund's coupling $J$ = 0.8 eV,
which were shown to describe Ca$_{1-x}$La$_{x}$FeAs$_{2}$ compounds \cite{Jiang16,cal}.
The temperature is set to $T$ = 116 K.
The experimental crystal structure determined by X-ray diffraction \cite{Katayama13} is used.
The virtual crystal approximation is adopted to investigate the electronic structure of the La doped compounds.

\begin{figure}[t]
\includegraphics[width=0.7\columnwidth]{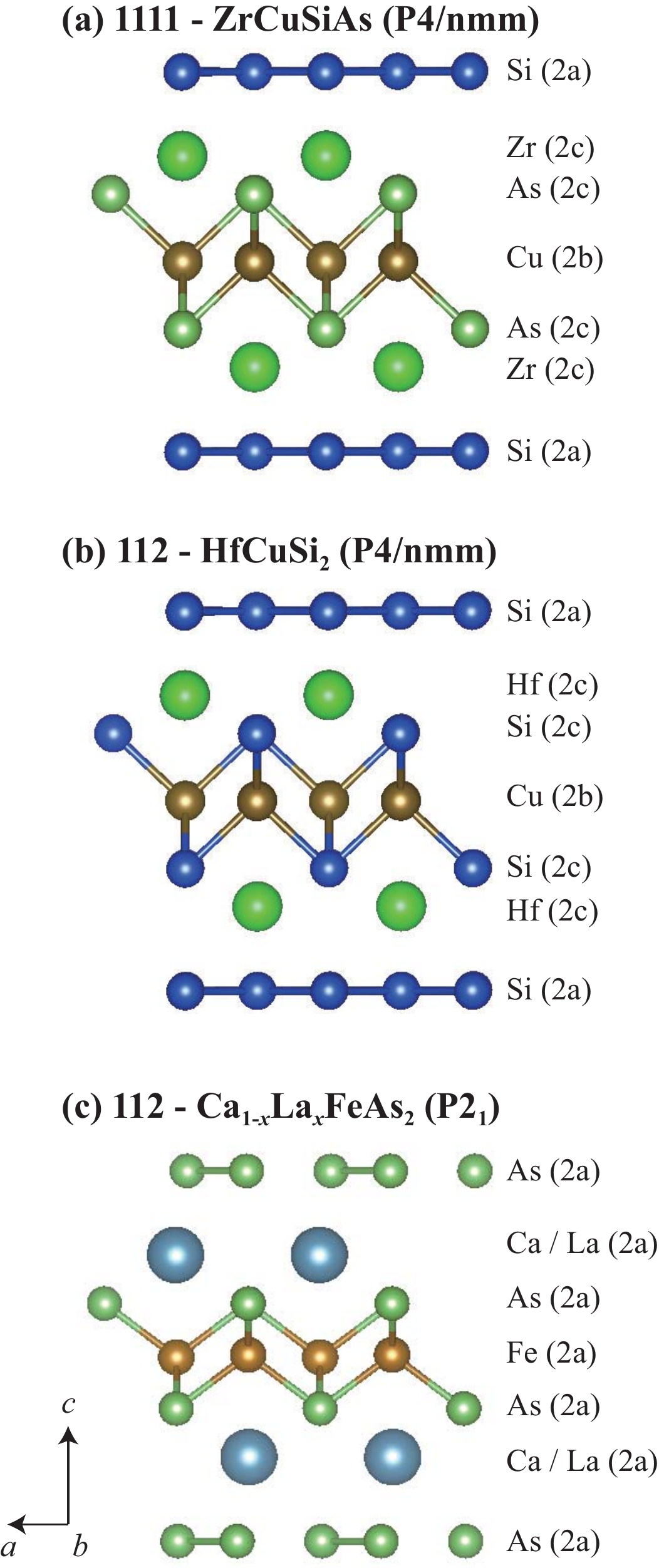}
\caption{Crystal structures related to iron 1111 and 112 phase systems.
Crystal structures of (a) ZrCuSiAs ($P4/nmm$), (b) HfCuSi$_{2}$ ($P4/nmm$),
and (c) Ca$_{1-x}$La$_{x}$FeAs$_{2}$ ($P2_1$).}
\label{related-structure}
\end{figure}

\section{Crystal Structure of
C\MakeLowercase{a}F\MakeLowercase{e}A\MakeLowercase{s}$_2$
and Its Connection to Other Pnictides}

\begin{table*}[t]
\begin{tabular}{|c|c|c|c|c|c|}
\hline
& \textbf{Space Group} & \multicolumn{4}{c|}{\textbf{Ion - Wyckoff Position}} \\
\cline{1-6}
ZrCuSiAs		&P4/nmm (\#129)&Zr - 2c& Cu - 2b& Si - 2a& As - 2c\\
\hline
LaFePO		&P4/nmm (\#129)&La - 2c& Fe - 2b& O - 2a& P - 2c\\
\hline
HfCuSi$_2$	&P4/nmm (\#129)&Hf - 2c& Cu - 2b& Si - 2a& Si - 2c\\
\hline
Ca$_{1-x}$La$_{x}$FeAs$_2$	&P2$_1$ (\#4)&Ca/La - 2a& Fe - 2a& As - 2a& As - 2a\\
\hline
h-CaFeAs$_2$	&P4/nmm (\#129)&Ca - 2c& Fe - 2b& As - 2a& As - 2c\\
\hline
\end{tabular}
\caption{Structural information on iron 1111 and 112 phase systems.}
\label{table-structure}
\end{table*}

In this section, we explain the details of the crystal structure of Ca$_{1-x}$La$_x$FeAs$_2$
and its connection to the 1111 family of superconductors with the ZrCuSiAs-type structure (see Fig. \ref{related-structure} and Table \ref{table-structure}).
The ZrCuSiAs-type structure has the simple tetragonal space group $P4/nmm$ (number \#129)
with two formula units per primitive unit cell.
In the superconducting pnictides with this structure,
the Fe ion occupies the same Wyckoff position as Cu in ZrCuSiAs,
$2b$ (1/4, 3/4, 1/2) with no internal parameter, and forms a layer
with the pnictogen in the same Wyckoff position as As in ZrCuSiAs, $2c$ (1/4, 1/4, $z$).
The internal parameter $z$ determines the pnictogen height, which is shown
to be sensitively linked to superconducting $T_c$ in these compounds (Fig. \ref{fig:Tc}).
In this layer, the Fe cations form a square plane with pnictogen ions above and below.
The pnictogen ions form edge-sharing tetrahedra around the Fe ions.
The other two ions are a cation on the $2c$ position and an anion on the $2a$ (3/4, 1/4, 0) position.
While in certain compounds these two ions can form a tightly bound layer together,
depending on the particular atoms there may or may not be any significant bonding between them,
and this gives rise to a wide range of internal parameter $z$ for the cation on the $2c$ position.

Having the same anions on both $2c$ and the $2a$ positions
in the ZrCuSiAs-type structure gives rise to the HfCuSi$_2$-type structure.
Note that in this structure, while both anions (Si) are the same,
their environments are significantly different.
The $2a$ position has both atoms in one unit cell on the same layer
(with coordinates (3/4, 1/4, 0) and (1/4, 3/4, 0)),
but the $2c$ position corresponds to atoms on two separate layers
(with coordinates (1/4, 1/4, $z$) and (3/4, 3/4, -$z$)) with the Fe (Cu in HfCuSi$_2$) layer in between.
The anions on the $2a$ site form a dense square net,
and often have strong covalent bonds between themselves,
whereas the anions on the $2c$ site hybridize
with the transition metals on the $2b$ site.
Compounds with the HfCuSi$_2$-type structure has been studied extensively
(for example see \cite{Verbovytskyy12,Brylak95}),
and hypothetical BaFeAs$_2$ and BaFeSb$_2$ with this structure
has been proposed to be high $T_c$ superconductors if synthesized \cite{Shim09}.
The physics that emerges from this square net of covalently bound anions
has drawn particular attention in the literature.
It has been shown that this layer can support conventional superconductivity \cite{Mizoguchi12},
as well as Weyl semimetallic phases \cite{Borisenko15}.
The wide bands that are formed from the $p$ orbitals of the main-block elements
on the square net also commonly give rise to Peierls type instabilities
that lead to the distortions of the crystal structure \cite{Papoian00},
which have been studied in detail in various compounds with a similar structure, including various arsenides
\cite{mozharivsky2001, mozharivskyj2000, rutzinger2009b, rutzinger2009a}.
These studies show that the distortions on the As layers can be tuned by both doping and pressure, and
studying the similar distortion present in CaFeAs$_2$ and its effect on superconductivity can be important.

\subsection{Difference between the structure of Ca$_{1-x}$La$_x$FeAs$_2$ and the HfCuSi$_2$-type structure}

\begin{figure}[t]
\includegraphics[width=8.5 cm]{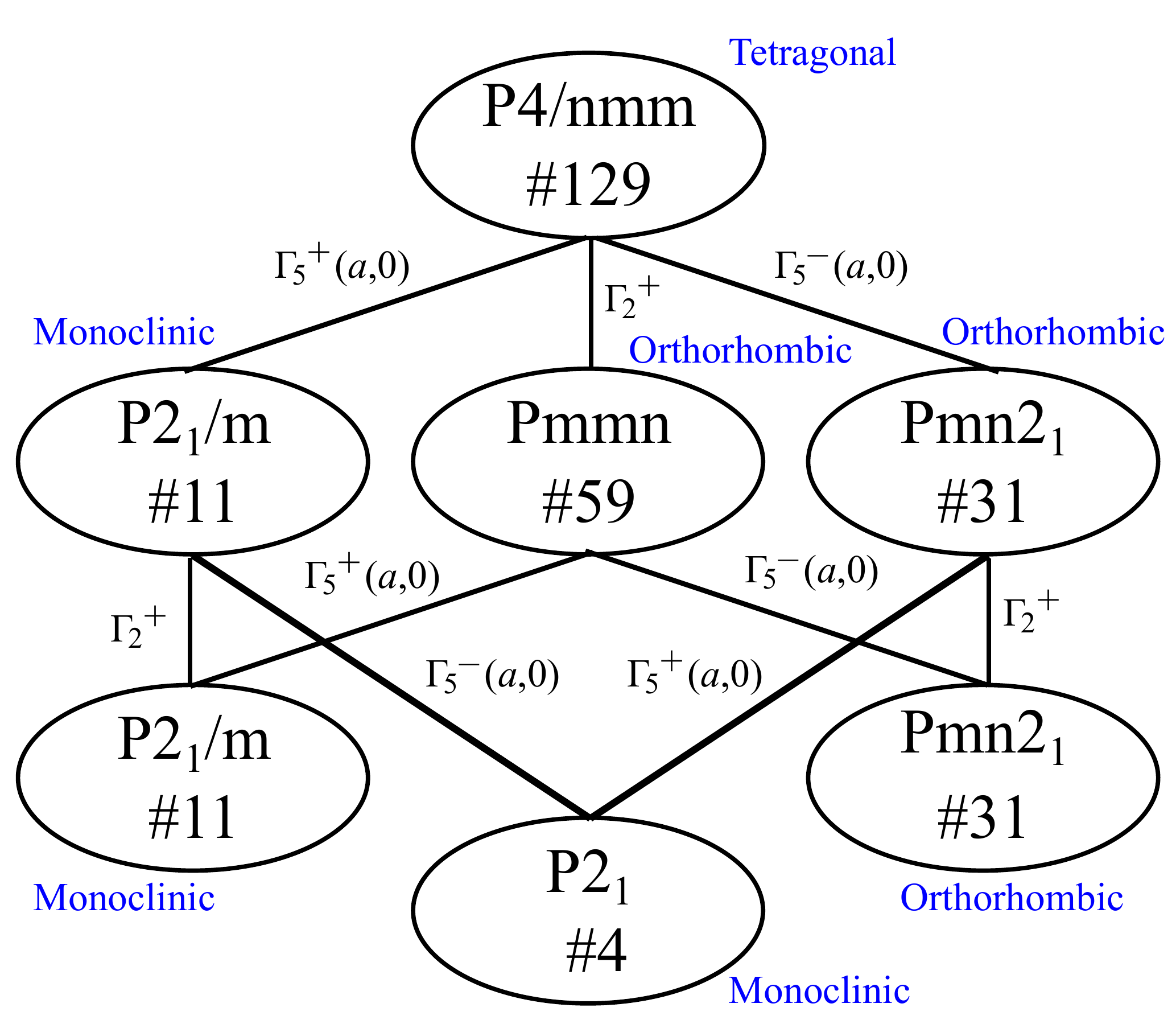}
\caption{(Color Online)
Group table showing the relevant structural phase transitions
between the tetragonal $P4/nmm$ and the monoclinic $P2_{1}$ structures.
The space group numbers are also shown below the Hermann-Mauguin notation.
Other possible structural phase transitions are also shown
with the corresponding symmetry of the lattice distortion.
}
\label{group}
\end{figure}

\begin{figure}[b]
\includegraphics[width=8.5 cm]{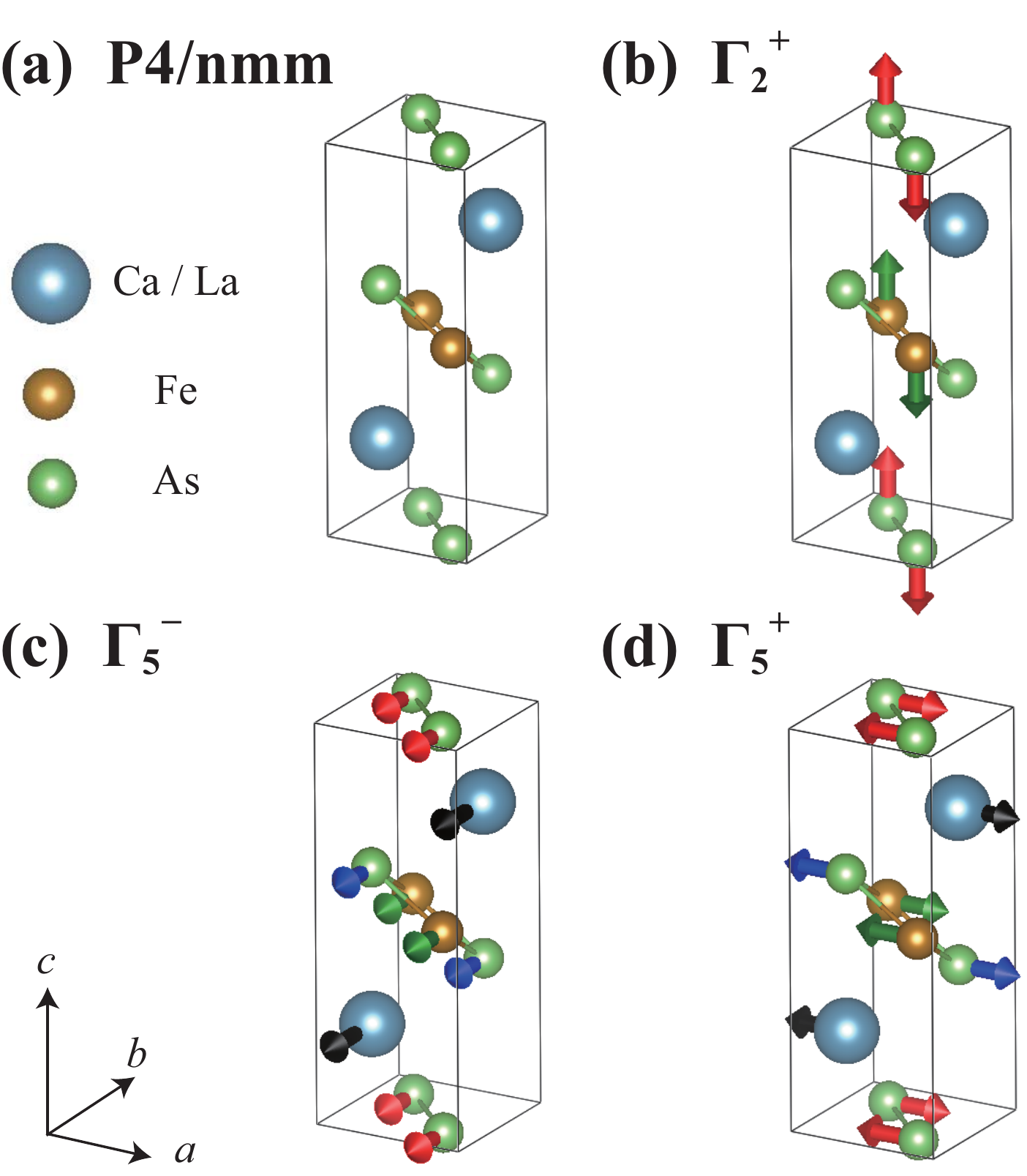}
\caption{(Color Online)
The sketch of the three different distortions present in the structure of Ca$_{1-x}$La$_x$FeAs$_2$ with respect to the
reference HfCuSi$_2$-type structure with space group \#129. (a) The structure sketched with the Fe layer in the center.
(b) The $\Gamma_2^+$ irrep involves a rumpling of the Fe and the As (which is in the CaAs layer) layers. (c) The $\Gamma_5^-$ irrep is the
in-plane polar irrep that involves a displacement of all the ions along the same direction. (d) The $\Gamma_5^+$ irrep
involves pair of ions moving opposite to each other in the same plane. This irrep is responsible of the formaion of the
As chains.
}
\label{fig:irreps}
\end{figure}

There are two very similar structures reported in the literature for lanthanide doped CaFeAs$_2$: A noncentrosymmetric structure with space group $P2_1$ (number \#4) and a centrosymmetric structure with space group $P2_1/m$ (number \#11).
(The former structure is reported, for example, for the La doped compound in Ref. \cite{Katayama13}, and the latter is reported, for example, for the Pr doped compound in Ref. \cite{Yakita14}. While second harmonic generation verifies the lack of inversion symmetry in the La doped compound \cite{Harter16}, no similar study exists for the Pr doped compound.)
There is a group-subgroup relationship between these two space groups,
and the only difference between the structures is a polar distortion,
discussed at the end of this subsection.

\begin{figure}[b]
\includegraphics[width=8.5 cm]{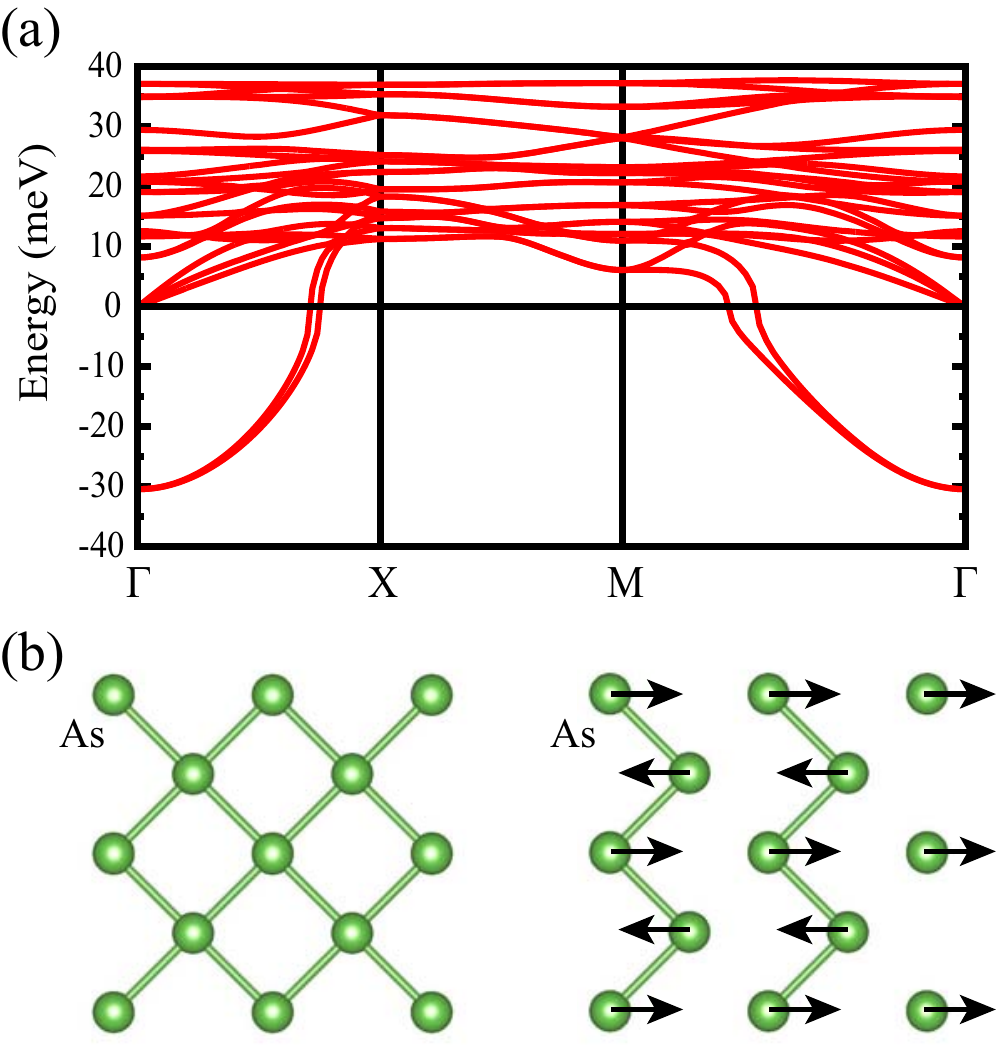}
\caption{(Color Online)
(a) Phonon dispersion curve
for the hypothetical tetragonal structure with $P4/nmm$ of CaFeAs$_{2}$.
The gamma-point phonon instability is clearly shown,
and the unstable phonon bands have the As square-net character.
(b) The corresponding unstable phonon normal mode.
The main lattice distortion occurs in the As square-net.
Black arrows represent the direction of the lattice distortion.
The As square-net is changed into the zig-zag As chain.
}
\label{phonon}
\end{figure}

The main feature that differentiates the structure
of Ca$_{1-x}$RE$_x$FeAs$_2$ from the HfCuSi$_2$-type structure
is a Peierls type distortion on the As square net as well (Fig.~\ref{phonon}).
This distortion is at the $\Gamma$ point of the unit cell
and transforms as the  $\Gamma_5^+$ irreducible representation
(irrep) (Fig.~\ref{fig:irreps} (d)).
(Throughout this study,
we label the irreps corresponding to structural distortions
using their labels for the space group \#129.)
This distortion form zig-zag chains from the As atoms,
and decreases the symmetry of the crystal significantly. (Even though this
distortion also displaces the Fe atoms to form Fe-Fe chains in a
similar fashion, and the Ca ions are also displaced, the amplitude of these
displacements, though nonzero by symmetry, are found to be very small in
the experimental structure and we hence ignore them.)
The $\Gamma_5^+$ distortion by itself breaks the four-fold rotational symmetry,
and renders Ca$_{1-x}$RE$_x$FeAs$_2$ unique in the sense that
it is the only pnictide superconductor with broken four-fold rotational symmetry
even above the nematic transition.
(The effect of this symmetry breaking (or lack thereof) on the nematic transition will be discussed latter.)

The structure that is obtained from the high symmetry reference \#129 by freezing
in the $\Gamma_5^+$ has the space group P2$_1$/m (\#11).
This symmetry is low enough that
a distortion that transforms as another irrep, $\Gamma_2^+$
(Fig. \ref{fig:irreps}(b)),
can also have a nonzero
amplitude without reducing the symmetry any further.
This irrep involves Fe and As (which is in the CaAs layer) layers' rumpling:
For example the two Fe ions in the same unit cell
are displaced by $\sim 0.01 \AA$ in opposite directions along the $c$ axis.
While in general it might be important that the Fe ions no longer form a perfect plane, this rumpling is so small that it can safely be ignored.
The small amplitude of this rumpling also
suggests that there is no driving force for the $\Gamma_2^+$ distortion,
but it is there only due to the
low symmetry induced by $\Gamma_5^+$,
very much like a secondary order parameter in a structural phase transition.

The structure that is obtained by the $\Gamma_2^+$ and $\Gamma_5^+$ distortions starting from the tetragonal HfCuSi$_2$-like reference structure is the centrosymmetric structure observed for some of the rare earth doped CaFeAs$_2$ compounds.
But, as mentioned before, some of these compounds actually have even lower symmetries.
The irrep that corresponds to the other structural distortion present in these structures is the in-plane polar irrep $\Gamma_5^-$ (Fig. \ref{fig:irreps}(c)).
This distortion is a overall displacement of all the
atoms parallel (or antiparallel) to each other
and breaks the inversion symmetry.
While, by symmetry,
every ion is displaced by this irrep, a symmetry mode amplitude analysis
of the experimental structure indicates that this displacement is more
than 80\% on the Ca layer, i.e. it is the Ca
(and the La atoms) that are displaced by far the most according to
this irrep. This is a very interesting observation
given the fact that Ca is by far the most electropositive element
in this compound and has a closed shell configuration,
and as a result, is not chemically active and should not be the driving force of a polar distortion. It is neither in a particular coordination geometry that is known to give rise to some type of geometric-ferroelectricity, which would be robust even though the compound is conducting \cite{benedek2016, ederer2006}.

Given the fact that this polar distortion is quite small on the Fe and As layers,
we do not analyse its effect on the electronic structure in detail,
and consider a complete study of the driving force behind it
beyond the scope of this work.
However, in passing,
we note that the evolutionary structure search we performed for CaFeAs$_2$,
discussed in the following sections,
does not predict a polar ground state.
Given the electropositivity of La,
it would be surprising if it was the driving ion for polarization in this compound.
The only possibility left seems to be that
the ordering of the La ions in the Ca layer gives rise to the polar distortion. This possibility is consistent with
the fact that the ionic radius of Pr is closer to Ca than La and therefore it is less likely to cause cation ordering
when substituting for Ca, however, these differences are all very small and it is not possible to go beyond the
level of speculation at this point.

\begin{figure*}[t]
\includegraphics[width=16 cm]{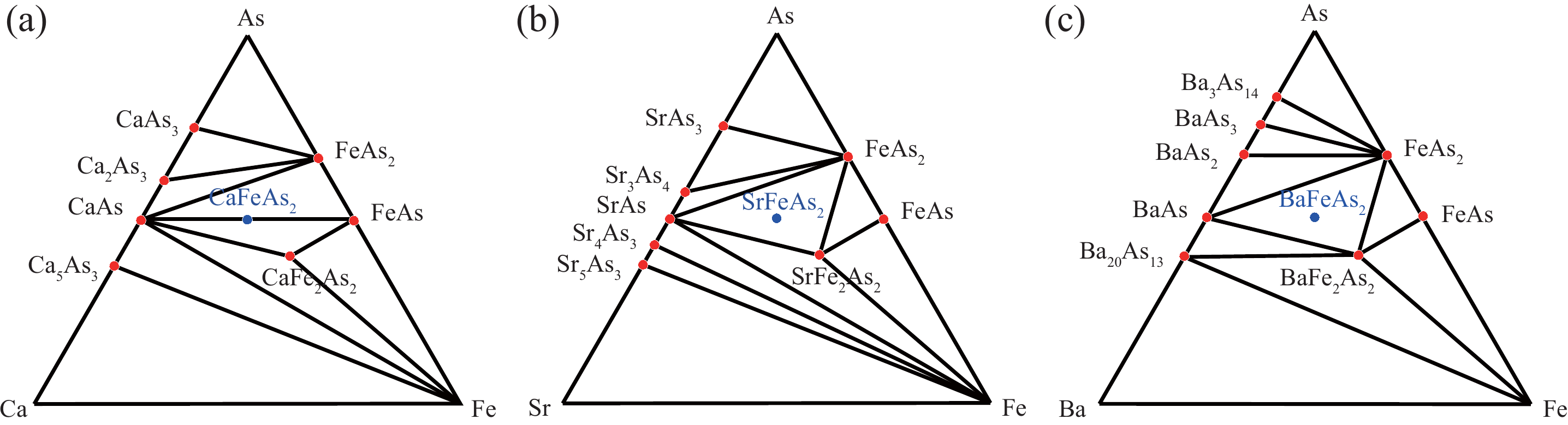}
\caption{(Color Online)
Ternary phase diagram for (a) CaFeAs$_{2}$,
(b) SrFeAs$_{2}$, and (c) BaFeAs$_{2}$.
Red and blue dots represent thermodynamically
stable and unstable phases, respectively.
CaFeAs$_{2}$, SrFeAs$_{2}$, and BaFeAs$_{2}$ are put
above the convex-hull,
and the energies above hull are 13 meV/atom,
24 meV/atom, and 17 meV/atom, respectively.
}
\label{ternary}
\end{figure*}

\subsection{Lack of tetragonal symmetry
and relation to electronic nematicity}

Figure \ref{group} shows various irreducible representations that
correspond to structural distortions from the tetragonal $P4/nmm$
structure of Ca$_{1-x}$La$_{x}$FeAs$_2$.
Even though there are orthorhombic phases listed in this figure,
these phases are different from (i.e. they belong to different space groups)
the orthorhombic phases 
realized below the structural transitions in, for example, LaFeAsO,
where the transition is from the tetragonal $P4/nmm$
to the orthorhombic $Cmma$ (\# 67) structure \cite{Nomura08}.
Similarly, in the 122 compounds that have a tetragonal structure with the ThCr$_{2}$Si$_{2}$ type
(space group $Fmmm$) at high temperature,
an electronic nematic phase breaks the four-fold rotational symmetry
and hence induces the structural transition
to an orthorhombic space group \cite{Chu10,Chu12,Kasahara12}.
In the case of Ca$_{1-x}$La$_{x}$FeAs$_{2}$,
the monoclinic structure is stable up to at least 450 K \cite{Katayama13}, and
no transition to the tetragonal phase is reported.
Given the one order of magnitude lower critical temperature of the nematic transition in
other iron pnictides, the monoclinic phase in CaFeAs$_{2}$ certainly does
not have a nematic origin.
As we discussed before, it is most likely due to a Peierls-like mechanism, active on the As layer
that forms the zig-zag chains.

Ca$_{1-x}$La$_{x}$FeAs$_{2}$ also undergoes a structural transition near its
antiferromagnetic ordering temperature \cite{Jiang16}. This transition is from monoclinic to
triclinic symmetry. However, the argument for electronic nematicity is not straightforward for this
transition either because the four-fold symmetry is already broken above this transition, in the monoclinic
structure.
We address this problem in the final section and show that the free energy expression for
CaFeAs$_2$ does not contain any terms that can give rise to a different character of the nematic transition than
the other iron pnictide superconductors. This justifies the studies such as references
[\onlinecite{Kawasaki15,Zhou16,Jiang16-2}] which compare the phase diagram of this compound with other iron pnictide compounds.

\section{First Principles Results}

\subsection{Crystal Structure of CaFeAs$_2$}

The hypothetical, tetragonal high-symmetry
phase of CaFeAs$_{2}$ has the $P4/nmm$ symmetry (Fig.~\ref{struct}(a)),
and it is the same structure as the one proposed by Shim \emph{et al.} \cite{Shim09}
for the Ba compounds with the same stoichiometry.
In order to check the stability of this structure with respect to
structural transitions, we calculated its phonon frequencies with DFT.
We present the resulting phonon dispersion curves in Figure \ref{phonon}(a).
There is a single, but two-fold degenerate $\Gamma$-point unstable phonon, which
is the \textit{chain-forming} instability that transforms as $\Gamma_5^+$.
There is no unstable $\Gamma_2^+$ mode, consistent with our claim in the previous
section that this irrep exists in the lower symmetry structure only because
of its coupling with the $\Gamma_5^+$ and not because there is a separate driving force
for this distortion.

The evolutionary structure prediction performed by USPEX predicts the lowest energy
crystal structure to have the $P2_{1}/m$ symmetry (Fig.~\ref{struct}(b)). The predicted
lattice constants are $a = 3.962 {\AA}$, $b = 3.896 {\AA}$, $c = 10.057 {\AA}$,
and $\beta = 91.135 ^{\circ}$ ($\alpha = \gamma = 90 ^{\circ}$).
This result is based on the spin non-polarized GGA functional.
Spin polarized GGA and GGA + $U$ ($U$ = 2, 4 eV)
schemes combined with USPEX give the same space group $P2_{1}/m$
with the similar (but, of course, different) lattice constants.
Note that the existence of the zig-zag As chains is also captured by the
evolutionary structure search, but the predicted structure is
centrosymmetric, both in line with the phonon calculations.

The experimental structural data for the La-doped compound (Ca$_{1-x}$La$_{x}$FeAs$_{2}$)
measured by X-ray diffraction has the monoclinic $P2_{1}$ symmetry with the lattice constants are
$a = 3.94710 {\AA}$, $b = 3.87240 {\AA}$, $c = 10.3210 {\AA}$,
and $\beta = 91.415 ^{\circ}$ ($\alpha = \gamma = 90 ^{\circ}$) \cite{Katayama13}.
The other X-ray diffraction experiment on Pr-doped compounds
(Ca$_{1-x}$Pr$_{x}$FeAs$_{2}$) were reported to have
the monoclinic structure with $P2_{1}/m$
having the inversion symmetry \cite{Yakita14}.
(The measured lattice constants are
$a = 3.9163 {\AA}$, $b = 3.8953 {\AA}$, $c = 10.311 {\AA}$,
and $\beta = 90.788 ^{\circ}$.)
The strong optical second-harmonic response
was recently observed in Ca$_{1-x}$La$_{x}$FeAs$_{2}$,
clearly implying that the crystal does not have the inversion symmetry \cite{Harter16},
but similar data does not exist for Ca$_{1-x}$Pr$_{x}$FeAs$_{2}$ to the best of our knowledge.
Hence, the first principles predicted structure is quite similar to
the experimental structure, except that it does not capture the possible
inversion symmetry breaking.

\subsection{Thermodynamic stability and convex hull construction}
Even though the methods we used so far can predict what the
crystal structure of CaFeAs$_2$ will be \textit{if it forms}, they do not
address the possibility of the constituent elements phase separating into different
compounds. In order to check the phase stability of CaFeAs$_2$,
we build the convex-hull for all known binary and ternary
compounds formed by these elements.
The convex-hull construction evaluates
the stability of a given compound
against any linear combination of possible compounds effectively
\cite{Ong08,Ong10,Longo14,Hoang15,Richards16}.
Hence, it is possible to determine whether
a given compound is stable or prefers to decompose to other compounds
(within the accuracy of DFT and
the approximations to the exchange correlation functional).
We show in Fig.~\ref{ternary}(a) the Ca-Fe-As phase diagram constructed from
the calculated GGA(PBE) total energy of all relevant phases
listed in the materials database \cite{mater_proj}.
Since the additional symmetry breaking due to magnetic order
(for example, antiferromagnetic order) is not considered
in the structural prediction performed by USPEX,
we take into account stripe magnetic order
in the predicted monoclinic $P2_{1}/m$ structure of CaFeAs$_{2}$.
Due to the stripe magnetic order, the monoclinic $P2_{1}/m$ structure
is further relaxed into a triclinic $P\bar{1}$ structure having lower symmetry.
The magnetic moment is 1.95 $\mu_{B}$/Fe and the total energy
is lower by 19.50 meV/atom.
CaFeAs$_{2}$ with the stripe magnetic order
is computed to be 13 meV/atom above the convex hull,
that is, it has a decomposition energy of 13 meV/atom
to CaAs and FeAs phases.
Reference \cite{Hautier12} has systematically
studied the error of DFT in similar predictions, and found that the errors can
be modeled by a normal distribution with a mean close to zero and a
standard deviation of 24 meV/atom.
Using this error bar, our calculation indicates that
the undoped CaFeAs$_{2}$ compound is, within the error bar, on the stability boundary.
It is possible that doping it with rare earth ions helps its stability.
%

Apart from a possible energetic gain, an effect of La doping would be an entropic gain
if there is no ordering of the La atoms. (A complete ordering of La is unlikely
because of its similar ionic radius to Ca.) The corresponding free energy is
$-T S = -k_{B}T \ln N$,
where $T$, $S$, $k_{B}$, and $N$ are temperature, entropy, Boltzmann constant,
and configuration number, respectively.
For example,
in order to simulate the phase stability of 25\% La-doped compounds,
we used a 2 $\times$ 2 $\times$ 1 supercell.
This supercell contains 8 Ca atoms, and
there are $\binom{8}{2} = 28$ configurations for substituting 2 La atoms for Ca.
Hence, we set the configuration number $N$ as 28 and
use $T$ = 1000 K, which is an estimate of synthesis temperature \cite{sala2014}.
The resulting entropy contribution for the La-doped compound is about $-9$ meV/atom.
This could render the compound marginally stable according to our calculations, however,
a more detailed study is necessary to address the energetics of doping in this compound.

\subsection{Hypothetical SrFeAs$_2$ and BaFeAs$_2$ compounds}

\begin{figure}[t]
\includegraphics[width=8.5 cm]{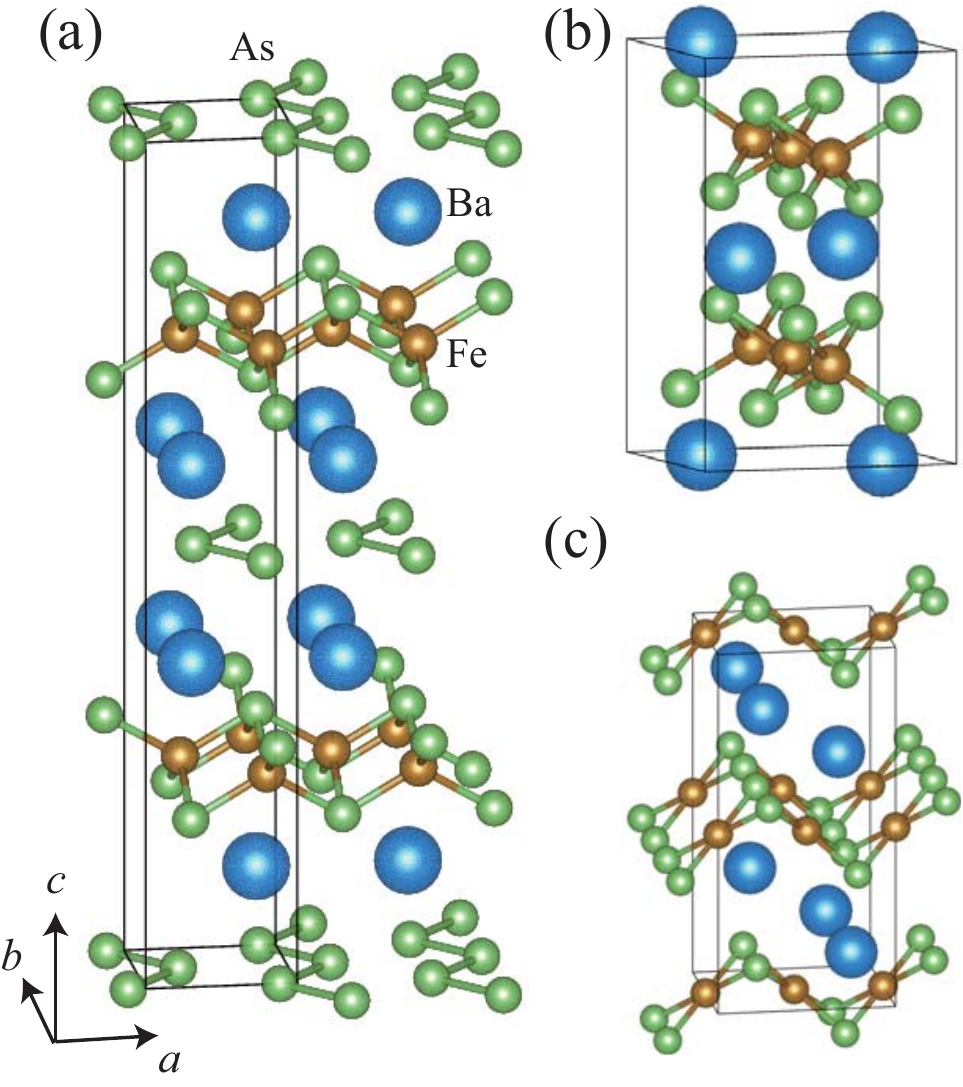}
\caption{(Color Online)
Crystal structures of BaFeAs$_{2}$ having the lowest energy
based on (a) spin non-polarized GGA,
(b) spin polarized GGA,
and (c) GGA + $U$ ($U$ = 2, 4 eV) schemes.
All of them are orthorhombic and
their space groups are $Imm2$ (\# 44),
$Cmma$ (\# 67), and $Cmcm$ (\# 63), respectively.
}
\label{Ba-struct}
\end{figure}

\begin{figure*}[t]
\includegraphics[width=16 cm]{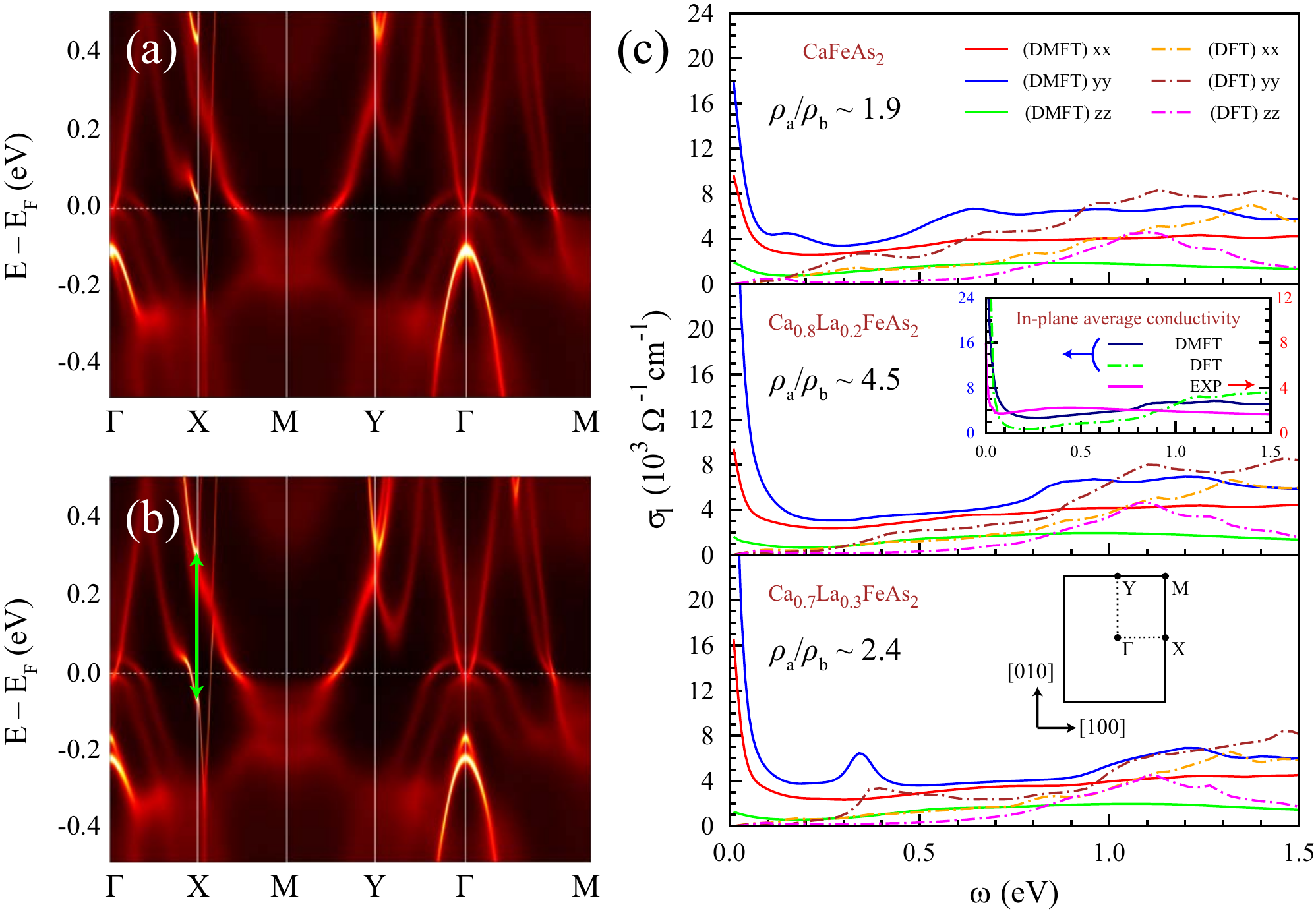}
\caption{(Color Online)
(a) $A(\textbf{k},\omega)$ of Ca$_{0.8}$La$_{0.2}$FeAs$_{2}$ at $T = 116 K$ as computed by DFT + eDMFT.
The Brillouin zone is shown in inset in the bottom of (c).
(b) $A(\textbf{k},\omega)$ of Ca$_{0.7}$La$_{0.3}$FeAs$_{2}$ at $T = 116 K$ as computed by DFT + eDMFT.
(The $A(\textbf{k},\omega)$ of the 30\% doped compound is also presented in Ref. \cite{Jiang16} with orbital
projection onto the in-plane $p$ orbitals of As ions forming the chains.)
(c) Optical conductivities within DFT + eDMFT method of CaFeAs$_{2}$ (top),
Ca$_{0.8}$La$_{0.2}$FeAs$_{2}$ (middle),
and Ca$_{0.7}$La$_{0.3}$FeAs$_{2}$ (bottom).
Optical conductivities (interband contributions only) within DFT method are also shown for comparison. (Intraband transitions within DFT give a delta function-like contribution at the zero frequency.)
The values of in-plane resistivity anisotropy $\rho_{a}/\rho_{b}$
correspond to the eDMFT results.
The in-plane average conductivity is shown in inset
for comparison between eDMFT, DFT, and experiment.
We used 0.01 eV for the broadening of intraband contributions in DFT.
Experiment data is for Ca$_{0.77}$Nd$_{0.23}$FeAs$_{2}$ at $T$ = 125 K,
and digitized from Ref. \cite{Yang16}.
For Ca$_{0.7}$La$_{0.3}$FeAs$_{2}$,
there is a low-energy peak of 0.34 eV in the optical conductivity,
which corresponds to the transition marked by the green arrow in (b).
}
\label{dmft}
\end{figure*}

%
To the best of our knowledge, there exists no report of the
synthesis of the Sr or Ba variants of CaFeAs$_2$ even with doping.
This is surprising given that Sr usually easily substitutes
for Ca, a fact that is supported by data-mining studies of existing
crystal structures \cite{hautier2010}.
In order to see if it is possible, in theory, to synthesize these compounds,
and whether they would have a crystal structure that is favorable for
pnictide superconductivity,
we also investigate the ground state structure for Sr and Ba compounds
using the evolutionary structure search method with several schemes for
the DFT part, such as spin non-polarized GGA,
spin polarized GGA, and GGA + $U$ schemes.
For the Sr compound, the lowest energy crystal structure
for the spin non-polarized GGA functional
is the monoclinic structure with the space group $P2_{1}/m$,
which is the same as the Ca compound.
This monoclinic structure with $P2_{1}/m$ is robust
with other spin polarized GGA and GGA + $U$ schemes.
Considering the stripe magnetic order
in the monoclinic $P2_{1}/m$ structure of SrFeAs$_{2}$,
the triclinic $P\bar{1}$ structure with the magnetic moment of 2.09 $\mu_{B}$/Fe
is obtained. The total electronic energy is lower by 23.08 meV/atom
compared to the monoclinic $P2_{1}/m$ structure
with no magnetic order (the nonmagnetic state).
As shown in Fig.~\ref{ternary}(b), SrFeAs$_{2}$ with the stripe magnetic order
is unstable with 24 meV/atom above the convex hull.
This energy above the convex hull is higher than that of the CaFeAs$_2$ compound,
which might explain why the Sr compound is not synthesized so far.

The situation for the Ba compound is quite different, and different
choices for the DFT scheme gives different results for the preferred structure.
The evolutionary structure search
with the spin non-polarized GGA functional
gives the orthorhombic structure with the space group $Imm2$ (\# 44)
(Fig. \ref{Ba-struct}(a))
as the ground state structure.
The obtained lattice constants are
$a = 4.059 {\AA}$, $b = 3.984 {\AA}$, $c = 23.161 {\AA}$,
and two FeAs layers and As zig-zag chains are contained in the unit cell.
This orthorhombic structure does not have inversion symmetry.
Evolutionary search using the spin polarized GGA functional
gives a structure that is orthorhombic
and has the space group $Cmma$ (\# 67).
Its lattice constants are
$a = 6.617 {\AA}$, $b = 5.977 {\AA}$, $c = 10.316 {\AA}$
(Fig. \ref{Ba-struct}(b)).
In this orthorhombic structure,
Fe atom is surrounded by As anions tetrahedrally,
however these tetrahedra form chains instead of a 2D FeAs layer,
which is not a structure that would favor high T$_c$ superconductivity
even if it actually formed.
Finally, the evolutionary structure search with GGA + $U$ ($U$ = 2, 4 eV)
gives the orthorhombic structure with space group $Cmcm$ (\# 63)
(Fig. \ref{Ba-struct}(c))
as the lowest energy structure.
The lattice constants of this structure are
$a = 6.086 {\AA}$, $b = 5.501 {\AA}$, $c = 12.621 {\AA}$.
Each Fe ion is again coordinated by 4 As ions, but the coordination
geometry is square rather than a tetrahedron. This is not a commonly
observed geometry for Fe, and this structure, even if it is synthesized,
would surely not favor high temperature superconductivity.

We consider the various possible magnetic orderings
among several predicted structures in BaFeAs$_{2}$
and calculate the total electronic energies.
Among them, the antiferromagnetic order
(with the magnetic moment of 2.17 $\mu_{B}$/Fe)
in the orthorhombic $Imm2$ (\#44)
has the lowest energy.
This antiferromagnetic order has the spins aligned the same
way as a single stripe magnetic order
within the $ab$ plane,
but they are ferromagnetic along the $c$ axis.
Therefore, this order is different
from the single stripe magnetic order exhibited in most iron pnictide superconductors,
where the spins are ordered in an antiferromagnetic fashion along the $c$-axis as well.
The inter FeAs layer distance in the orthorhombic $Imm2$ is 11.55 \AA,
so that the quite large inter-layer distance might affect the inter-layer magnetic ordering.

We construct the convex hull for BaFeAs$_{2}$ compound with the stripe magnetic order.
As shown in Fig. \ref{ternary}(c), BaFeAs$_{2}$ compound is unstable
with 17 meV/atom above the convex hull.
The energy above the hull is somewhat higher than that of CaFeAs$_{2}$, however,
it is quite reduced compared to that of SrFeAs$_{2}$.
We predict that the rare-earth doped BaFeAs$_{2}$ might be possible to synthesize
considering that the rare-earth doping makes Fe 112 phase be more stable
(for example, the energetic gain from entropy as discussed before).
From Fig. \ref{ternary}, we can get some information
about mixtures of the essential compounds to grow the target material.
Since the reaction CaAs + FeAs $\rightarrow$ CaFeAs$_{2}$ has
the minimum enthalpy of formation of 13 meV/atom (Fig. \ref{ternary}(a))
and the rare-earth doping makes this enthalpy of formation be lower,
heating (supplying energy) a mixture of CaAs, FeAs, and rare-earth compounds
is essential to grow the rare-earth doped CaFeAs$_{2}$ compound \cite{sala2014,Yakita14}.
On the other hand, BaFeAs$_{2}$ compound would decompose into
BaAs, FeAs$_{2}$, and BaFe$_{2}$As$_{2}$.

\begin{figure}[t]
\includegraphics[width=7 cm]{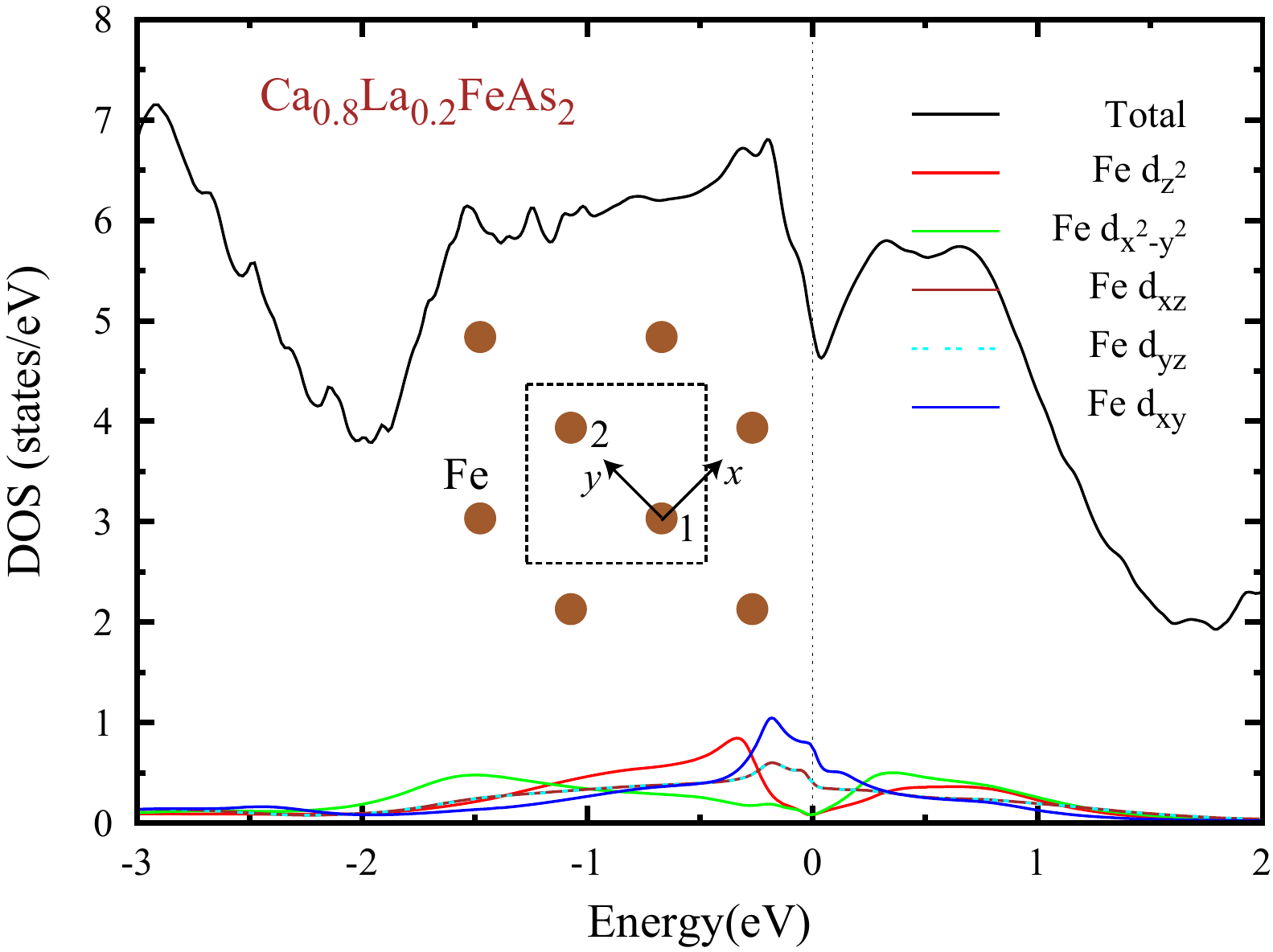}
\caption{(Color Online)
DOS of Ca$_{0.8}$La$_{0.2}$FeAs$_{2}$.
Total and projected DOS of Fe $3d$ orbitals
calculated by DFT + eDMFT at $T$ = 116 K.
The inset shows the direction of local axes
for the projected DOS of Fe $3d$ orbitals.
Our choice of Cartesian axes are 45 degrees rotated
with respect to the $\langle 100\rangle$ axes.
The occupations of Fe $d_{xz}$ and $d_{yz}$ are same,
giving the zero orbital polarization.
}
\label{dos}
\end{figure}

\begin{figure}[t]
\includegraphics[width=8.5 cm]{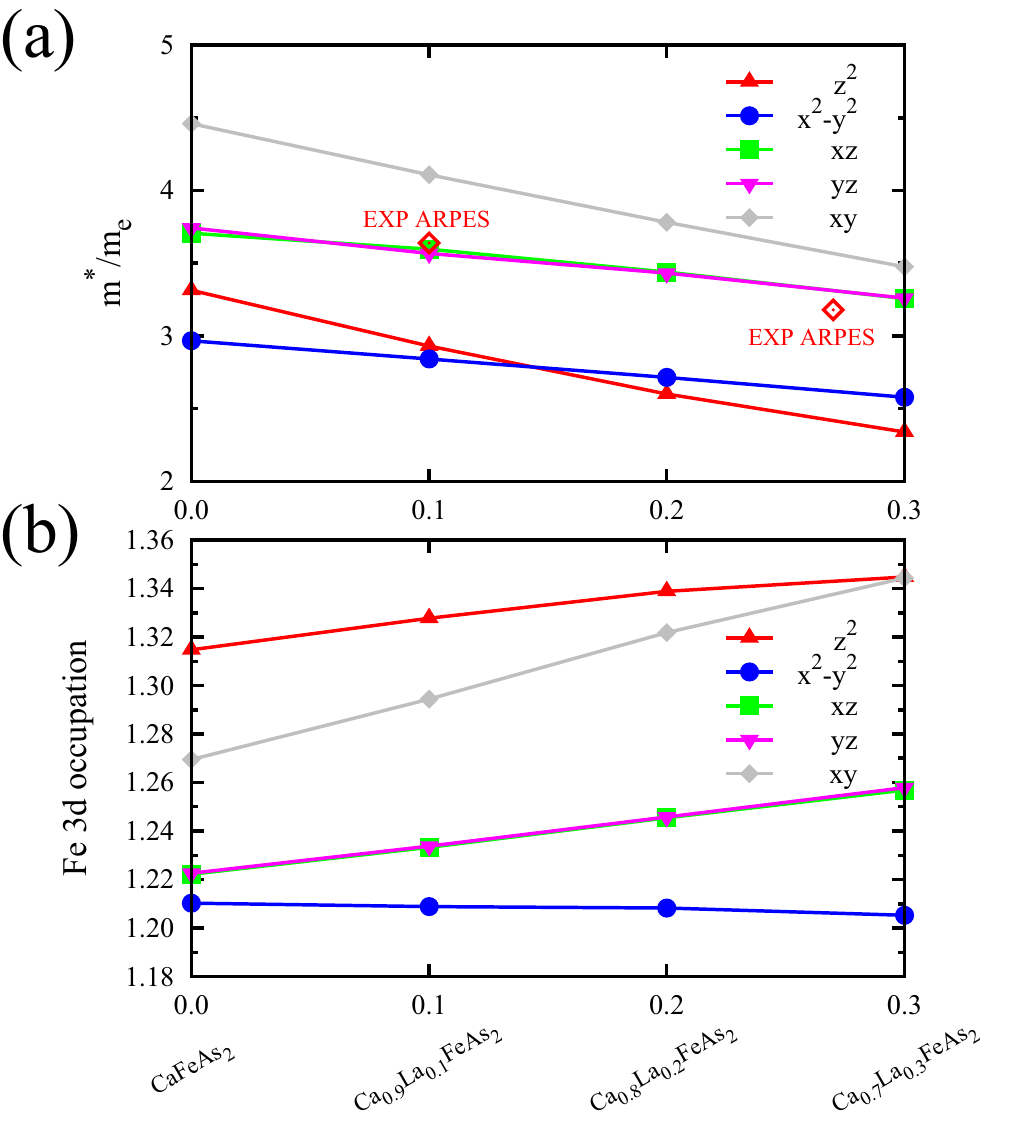}
\caption{(Color Online)
(a) The mass enhancement $m^{*}/m_{e}$
of the iron $3d$ orbitals upon doping
calculated by the DFT + eDMFT.
The experimental data (open diamond) are obtained from
angle-resolved photoemission spectroscopy experiments
in Refs. \cite{Li15,Jiang16}.
(b) The orbital occupation of the iron $3d$ orbitals
upon doping.
}
\label{mass}
\end{figure}

\subsection{Electronic structure of Ca$_{1-x}$La$_{x}$FeAs$_2$}

Figure \ref{dmft}(a) shows the momentum-resolved electronic spectral function
$A(\textbf{k},\omega)$ of Ca$_{0.8}$La$_{0.2}$FeAs$_{2}$
at $T$ = 116 K as computed by DFT + eDMFT.
A fast-dispersing band near the X point
has the dominant character of the zig-zag As chain \cite{Jiang16,Li15},
and this band goes down below the Fermi level upon doping as shown in Fig.~\ref{dmft}(b),
indicating that the spacer of the zig-zag As chain
has a crucial role in doping \cite{Jiang16}.
The electronic anisotropy is evident in the band structures,
especially along $\Gamma$-X and along $\Gamma$-Y directions,
as shown in Figs.~\ref{dmft}(a) and (b).
In order to clarify the origin of the anisotropy,
we calculated the orbital polarization
between the Fe $d_{xz}$ and $d_{yz}$ orbitals,
$\phi = (n_{xz}-n_{yz})$,
where $n_{xz}$($n_{yz}$) denotes
the occupation of the Fe $d_{xz}$($d_{yz}$) orbital.
$\phi$ is zero for all compositions at given temperature $T$ = 116 K
(Fig.~\ref{dos}).
Therefore, the anisotropy does not originate
from the electronic nematic phase
and it is due to the structural anisotropy arising from the zig-zag As chain.

This anisotropy is also reflected in the optical conductivity.
Figure \ref{dmft}(c) illustrates the doping dependence in the optical conductivity.
Comparing the optical conductivity of CaFeAs$_{2}$
to that of Ca$_{0.8}$La$_{0.2}$FeAs$_{2}$,
the $xx$ and $zz$ components does not change much upon doping.
However, there is a significant change in the $yy$ component
upon doping, and the dc conductivity is enhanced significantly.
This results in a large in-plane resistivity anisotropy
as much as $\rho_{a}/\rho_{b} \approx 4.5$,
where $\rho_{a}$ and $\rho_{b}$ are dc resistivities
along $a$-axis ($x$-axis) and $b$-axis ($y$-axis), respectively.
The in-plane resistivity anisotropy has only a structural origin,
and it is different from a electronic nematicity driven anisotropy
which was found in Ba(Fe$_{1-x}$Co$_{x}$)$_{2}$As$_{2}$ \cite{Chu10}.
The metallic zig-zag As chain is formed along the $b$-axis,
so that it gives the higher conductivity along the $b$-axis.
Note that
the average in-plane optical conductivity for Ca$_{0.77}$Nd$_{0.23}$FeAs$_{2}$
was recently reported experimentally \cite{Yang16},
and it is consistent with our calculated optical conductivity
for Ca$_{0.8}$La$_{0.2}$FeAs$_{2}$
as shown in inset of Fig. \ref{dmft}(c).

For further La-doping (Ca$_{0.7}$La$_{0.3}$FeAs$_{2}$),
only the in-plane $xx$ component of the dc conductivity
changes significantly.
Besides, a low-energy peak of 0.34 eV appears
in the $yy$ component of the optical conductivity.
The low-energy peak of 0.34 eV corresponds to
the coherent interband transition within the zig-zag As chain,
which is marked by the green arrow in Fig.~\ref{dmft}(b).

Since the parent compound of Ca$_{1-x}$La$_{x}$FeAs$_{2}$
is regarded as Ca$_{0.7}$La$_{0.3}$FeAs$_{2}$ \cite{Jiang16},
the superconductivity in Ca$_{0.8}$La$_{0.2}$FeAs$_{2}$
arises from the hole doping through Ca substitution on the La sites.
The resistivity anisotropy is a nonmonotonic function of doping
as shown in Fig.~\ref{dmft}(c),
and it reaches a maximum near the superconducting dome \cite{Chu10}.

The plasma frequencies \cite{optics} $\omega_{p,xx}$, $\omega_{p,yy}$, and $\omega_{p,zz}$
in the $x$, $y$, and $z$ directions obtained by DFT calculations
are 2.47 (2.56), 3.36 (3.56), and 0.57 (0.52) eV, respectively, for Ca$_{0.8}$La$_{0.2}$FeAs$_{2}$ (Ca$_{0.7}$La$_{0.3}$FeAs$_{2}$).
The estimated in-plane resistivity anisotropy in DFT calculations
$\rho_{a}/\rho_{b} \simeq \omega_{p,yy}^2/\omega_{p,xx}^2$ is about
1.85 (1.94) for Ca$_{0.8}$La$_{0.2}$FeAs$_{2}$ (Ca$_{0.7}$La$_{0.3}$FeAs$_{2}$).
The in-plane resistivity anisotropy is diminished compared to
the eDMFT result giving the quite large in-plane resistivity anisotropy.
The anisotropy difference between DFT and eDMFT calculations
comes from the correlation effect on iron $3d$ orbitals.
The resistivity along $b$-axis ($\rho_{b}$) is almost dominated by
the non-correlated zig-zag As chain.
However, the resistivity along $a$-axis ($\rho_{a}$) is enhanced by the
correlated FeAs layer in the systems.
Therefore the correlation effect enhances the in-plane resistivity anisotropy
$\rho_{a}/\rho_{b}$
induced by the structural anisotropy exhibited in the systems.
Compared eDMFT with DFT optical conductivities,
the eDMFT calculations show higher spectral weights at low energy.
It is due to the incoherent spectral weight induced from the local correlation effect,
which is well described by eDMFT, however is not present in DFT.
The incoherent spectral weight is also clearly shown in inset of Fig. \ref{dmft}(c),
where we have compared the eDMFT and DFT calculations
with the measured in-plane average conductivity.
Based on the high spectral weight at low energy ($\sim$0.4 eV) in experiment,
eDMFT calculations give better description rather than DFT.
But still, eDMFT and experiment seem to have rather different positions of maxima,
at 1.2 eV and at 0.4 eV, respectively.
Note that when CaFeAs$_{2}$ is doped,
its crystal structure could change as well.
However, we do not take this effect into account,
and the change in the anisotropy we observe
is purely due to changes in the electronic structure,
and not due to changes in the crystal structure.


The mass enhancement $m^{*}/m_{e}$ of the iron $3d$ orbitals
is reduced upon doping as shown in Fig. \ref{mass}(a).
This means that the band becomes more coherent upon doping
and this leads to enhance the Drude peak in
the optical conductivity upon doping in Fig. \ref{dmft}(c).
Note that the mass enhancement is not equal in all orbitals,
and $t_{2g}$ obitals have larger enhancement than $e_{g}$ orbitals.
Among the $t_{2g}$ orbitals, the $xy$ orbital has the largest mass enhancement.
These behaviors are quite consistent with the previous eDMFT calculation
in iron-based superconductors
\cite{Yin11,Yin11np,Skornyakov10,Tomczak12,Ferber12}.
We also show the mass enhancement extracted from
angle-resolved photoemission spectroscopy (ARPES) experiments \cite{Li15,Jiang16}.
(With help of polarization-dependent ARPES experiments \cite{Li15},
it is possible to extract the band dispersion having iron $d_{xz}$ and $d_{yz}$ orbital characters
near the Fermi level and calculate the corresponding mass enhancement for individual orbitals.)
A good agreement between the eDMFT and experiment is shown in Fig. \ref{mass}(a).

Figure \ref{mass}(b) shows the orbital occupation
of iron $3d$ orbitals.
Upon doping the $x^2-y^2$ orbital has almost constant occupation,
and $xz/yz$ and $z^2$ orbitals have the similar increment in the orbital occupation.
(The increment from $x$ = 0 to $x$ = 0.3 compound
is 0.030 for $z^2$, and 0.035 for $xz/yz$ orbitals.)
The largest increment in the orbital occupation of iron $3d$ orbitals
is the $xy$ orbital
and is 0.075 from $x$ = 0 to $x$ = 0.3 compound.
Since this system has the metallic As spacer,
there is some additional charge
in the spacer As $4p$ orbitals upon doping.
The change in the charge on the As $4p$ orbitals between $x$ = 0 and $x$ = 0.3
is about 0.05.
This shows, again, that the metallic As spacer has an important role
in the doping process.

Note that two orbitals $z^2$ and $xy$ have very different mass enhancements
among other $3d$ orbitals
and their enhancements change a lot as a function of doping.
The orbital occupation of $z^2$ is the largest among other $3d$ orbitals
over the doping ratio up to 30 \%
and that of $xy$ is significantly increased upon doping.
These are effects beyond DFT and are very important for the resistivity anisotropy
realized in eDMFT calculations.

\section{Landau Free Energy}

In this section, we outline the basics of a Landau free energy relevant to Ca$_{1-x}$La$_{x}$FeAs$_{2}$,
as well as other ternary iron pnictide superconductors.
The reason for such a study
is that the absence of the $C_4$ symmetry in the high-temperature phase of
CaFeAs$_2$ might lead a misconception that the nematic transition in this compound
has to be significantly different from its counterpart in other
tetragonal iron pnictides. Below, we show that this is not the case. Even though there have been various studies which wrote down the Landau theory for these systems,
such as references \cite{baek2015, fernandes2012, cano2010},
to the best of our knowledge there are no studies which emphasize the difference between ferro- and antiferro-orbital orders and their connection with the nematicity.

\begin{figure}[t]
\includegraphics[width=0.75\columnwidth]{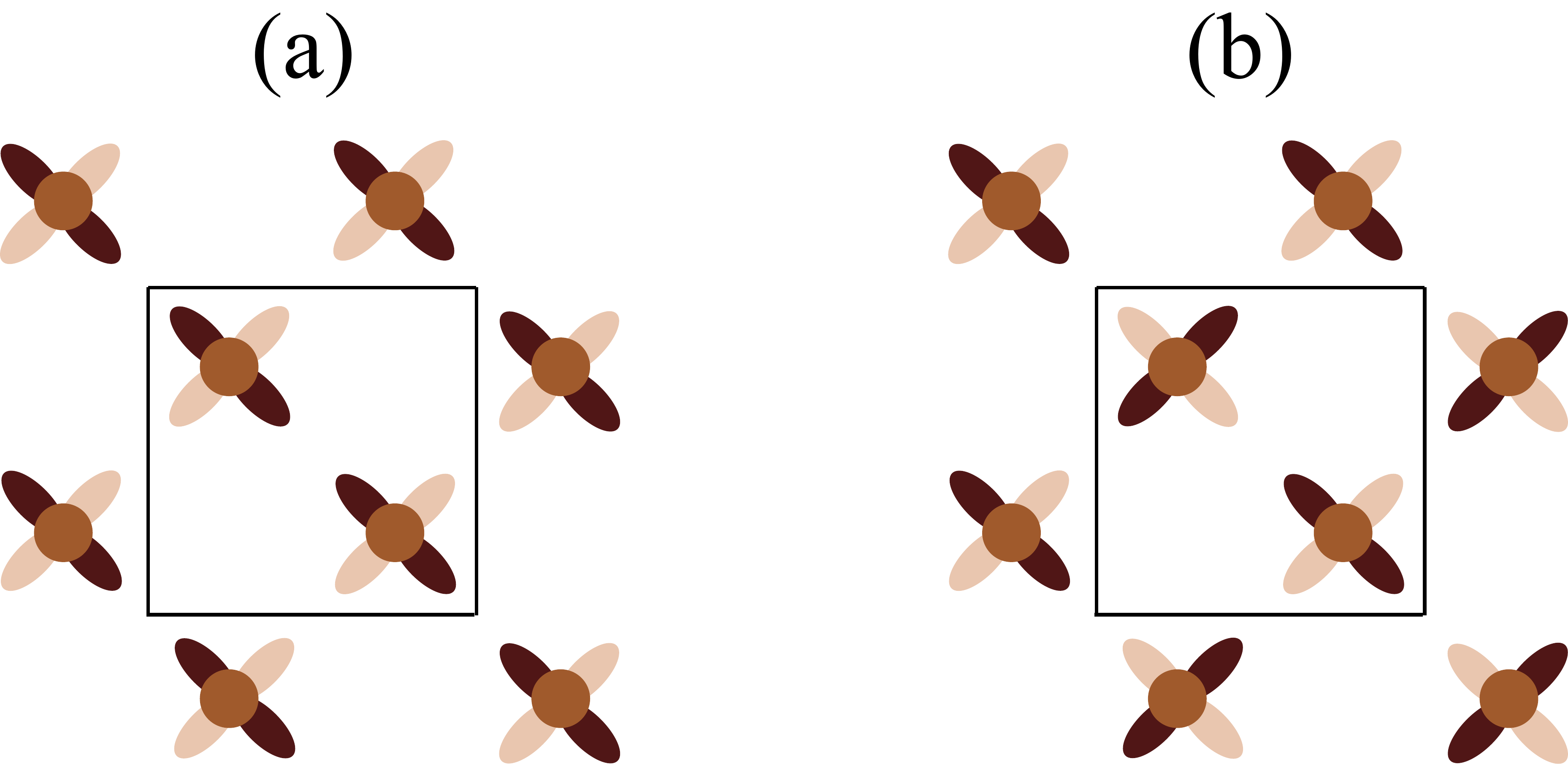}
\caption{(a) The ferro and (b) the antiferro orbital orders. }
\label{orbitalorder}
\end{figure}

We start by writing a Landau free energy expansion around
a high symmetry tetragonal phase of these compounds.
(A tetragonal phase only serves as a reference structure.)
The point we would like to emphasize is that
the primitive unit cell of this high symmetry phase contains two formula units.
As a result, even though many model tight-binding studies usually consider a single Fe atom,
we need to consider a two-Fe primitive cell (inset of Fig. \ref{dos})
when building a Landau theory.
We label the two Fe atoms with numbers 1 and 2,
and choose cartesian axes such that $x$ and $y$ directions point
towards nearest neighbors (inset of Fig. \ref{dos}).
We define the orbital polarization for each atom,
$\phi_i$ ($i$ = 1, 2), as the difference between
the occupations of $d_{xz}$ and $d_{yz}$ orbitals:
\begin{equation}
\phi_i = n_{i,xz} - n_{i,yz}.
\end{equation}
The two cases where the signs of the orbital polarization of the two atoms
are the same or opposite correspond to
the ferro- and antiferro-orbital orders (Fig. \ref{orbitalorder}).
We label the corresponding order parameters
by $\phi_+$ and $\phi_-$ as follows:
\begin{equation}
\phi_+ = \frac{\phi_1 + \phi_2}{2},
\end{equation}
\begin{equation}
\phi_- = \frac{\phi_1 - \phi_2}{2}.
\end{equation}

\begin{figure}[t]
\includegraphics[width=1.0\columnwidth]{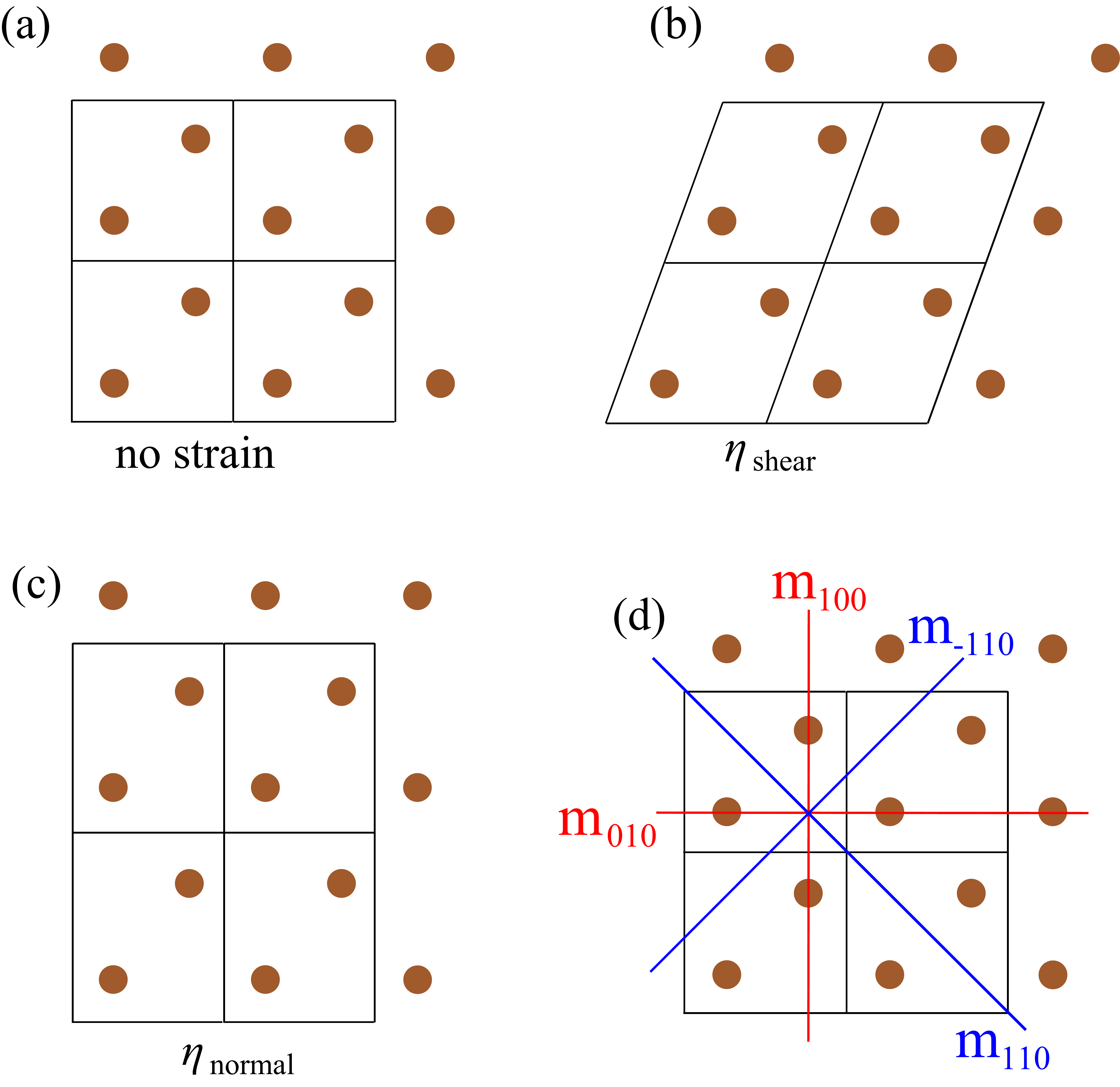}
\caption{Sketch of the Fe plane with
(a) no strain, (b) shear strain, and (c) normal strain.
While there are multiple components of possible shear and normal strains,
we only show the relevant ones,
the $\eta_{12}$ shear strain and $\eta_{11}$ normal strain. (d) In the nonstrained structure, there are four mirror planes that are not parallel to [001]. Normal strain breaks the mirror symmetry of only two of these planes
(\textbf{m$_{110}$} and \textbf{m$_{\bar{1}10}$}, shown in blue)
whereas the shear strain breaks only the other two
(\textbf{m$_{100}$} and \textbf{m$_{010}$}, shown in red).
}
\label{strain}
\end{figure}

While the onset of either $\phi_+$ or $\phi_-$ would break the symmetry
between $d_{xz}$ and $d_{yz}$ orbitals on an individual Fe ion,
these two order parameters break different space group symmetries of the crystal.
The nematic transition commonly observed in iron pnictides
involve the ferro orbital order $\phi_+$.
This order parameter breaks the four-fold rotational symmetry in a specific way
- it chooses one of the $x$ or $y$ cartesian axes over the other,
and hence can couple bilinearly with the shear strain,
which we denote as $\eta_s$ (Fig. \ref{strain}(b)),
so the free energy has a term $\sim\eta_s \phi_+$.

At this point, we would like to note that
the effect of the shear strain $\eta_s$ on the iron sublattice
is to convert it from a square to a tetragonal one.
As a result, for models considering a single-Fe unit cell,
the strain relevant to the nematic transition is a normal strain,
but not a shear strain.
However, the normal strain in the actual crystallographic unit cell
(Fig. \ref{strain}(c)) is totally different.
It is this normal strain $\eta_n$ that is present
at high temperature in Ca$_{1-x}$La$_{x}$FeAs$_{2}$.
While $\eta_n$ also breaks the four-fold rotational symmetry,
it does in a different way than $\eta_s$
and does not differentiate between $d_{xz}$ and $d_{yz}$ orbitals on the Fe atoms.
As a result, there is no bilinear coupling $\sim \eta_n \phi_+$
in the Landau free energy.

Another way to think about the shear and normal strains is to consider the mirror planes present in the reference structure. In the tetragonal structure of iron pnictides, there are 4 mirror planes that are perpendicular to the Fe layers (Fig. \ref{strain}(d)).
The presence of \textbf{m$_{100}$} and \textbf{m$_{010}$}
(shown in red in the figure) flips the $x$ and $y$ axes on a Fe atom
and as a result imposes the condition that $\phi_i=0$.
The normal strain breaks the \textbf{m$_{110}$} and \textbf{m$_{\bar{1}10}$} mirror symmetries but preserves \textbf{m$_{100}$} and \textbf{m$_{010}$}.
So, even though it breaks the $C_4$ symmetry, it does not create an orbital order.
The shear strain does the opposite, it breaks the \textbf{m$_{100}$} and \textbf{m$_{010}$} mirror symmetries but preserves \textbf{m$_{110}$} and \textbf{m$_{\bar{1}10}$}.
This gives rise to not only a nonzero $\phi_i$ but also a nonzero $\phi_+$.

\begin{figure}[t]
\includegraphics[width=.50\columnwidth]{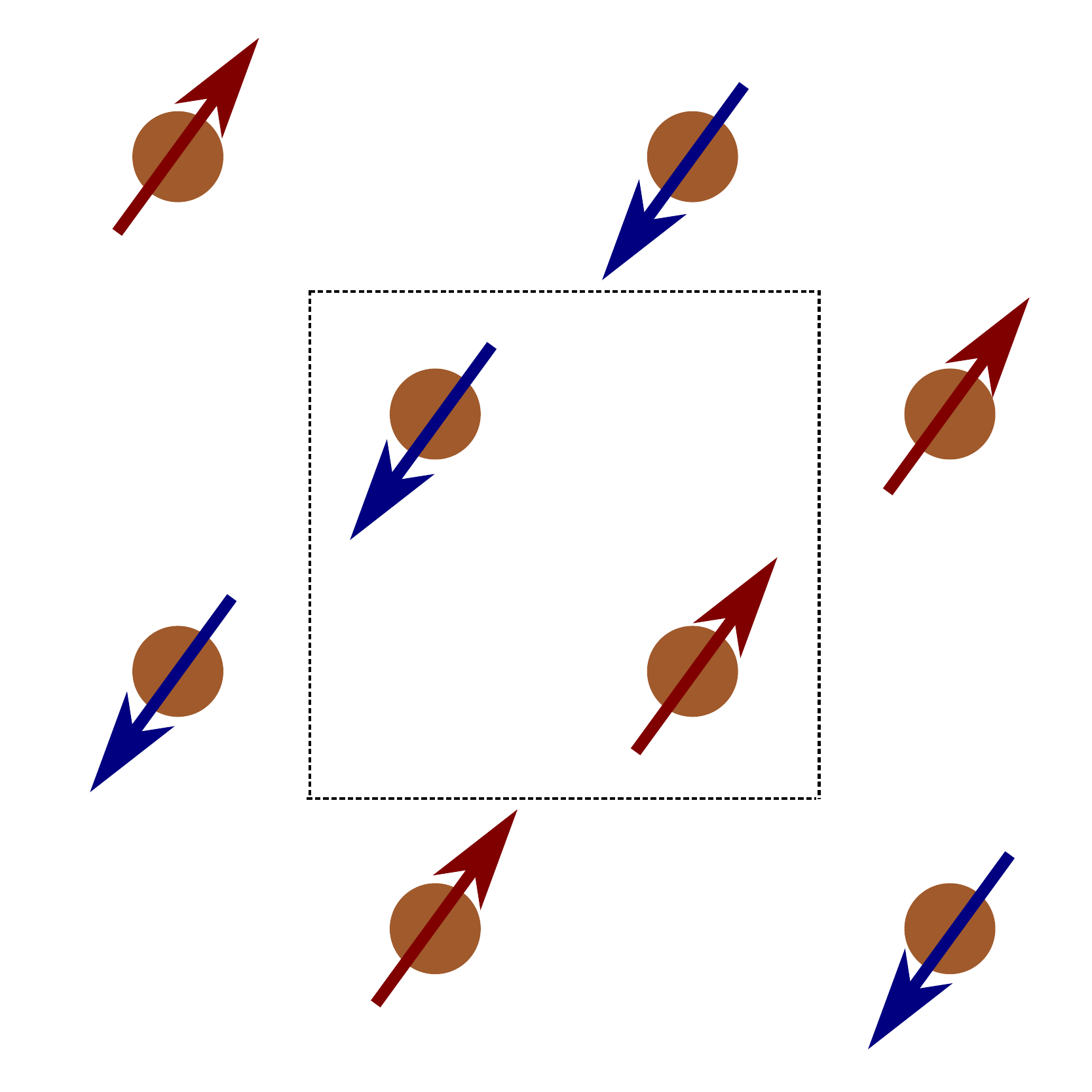}
\caption{The stripe AFM order with its wave vector
along the $[\bar{1}10]$ direction ($L_y$).
Note that in our theory we do not take into account the spin orbit coupling
and consider only collinear spin arrangements.
While the relative orientations of the spin moments are meaningful,
the direction that all the spins are parallel or antiparallel to is not.
The stripe AFM order breaks the \textbf{m$_{100}$}
and \textbf{m$_{010}$} mirror symmetries.
}
\label{AFM}
\end{figure}

The stripe antiferromagnetic (AFM) order
observed in iron pnictides has a $k$-vector
along the $\langle 110\rangle$ family of directions,
and involves a doubling of the unit cell.
We denote the AFM order parameters
with different wave vectors as $L_x$ and $L_y$ (Fig. \ref{AFM})
\footnote{Note that unlike many other studies where the site magnetic moments ($m_i=n_{i\uparrow}-n_{i\downarrow}$)
are introduced, we work directly with the AFM order parameters, since there is no other
magnetic order that is relevant to these compounds.}.
At the lowest order, there is no $\sim L_\alpha$ term in the free energy
since there is no other parameter that breaks the time reversal symmetry.
However, the onset of the AFM order breaks the four-fold rotational symmetry
and differentiates between $x$ and $y$ axes.
It also breaks the \textbf{m$_{100}$}
and \textbf{m$_{010}$} mirror symmetries.
(Whether the AFM order breaks the other mirror symmetries,
\textbf{m$_{110}$} and \textbf{m$_{\bar{1}10}$},
depends on the direction of the magnetic moments.
Even though the wave vector of the AFM order is preserved by
these two mirror operations, depending on the magnetic easy axis,
they might lead to a rotation of the magnetic moments,
and hence these mirror symmetries too might be broken by the onset of magnetic order.)
As a result, $L_\alpha$ can couple with $\eta_s$ at linear order
and it leads free energy to have terms that go as $\sim (L_x^2 - L_y^2) \eta_s$.
Similarly, terms such as $\sim (L_x^2 - L_y^2) \phi_+$ also exist.
However, the staggered orbital order $\phi_-$ does not couple
to any of these parameters at linear order,
and only biquadratic couplings such as $\sim (L_x^2+L_y^2)\phi_-^2$ exist,
since such biquadratic couplings exist between any two order parameters.

Gathering all these terms together,
we obtain the following free energy expression
valid for a high symmetry (tetragonal) pnictide,
that takes into account all the terms that couple to the ferro-orbital order:
\begin{widetext}
\begin{eqnarray}
\mathcal{F}&=& a_+ \phi_+^2 +b_+ \phi_+^4 +a_-\phi_-^2+b_-\phi_-^4 + c(\phi_+^2\phi_-^2) \\ \nonumber
   &&+d_{s+}(\eta_s \phi_+) + e_{s+}(\eta_s \phi_+^2)    +e_{n+}(\eta_n \phi_+^2) +e_{s-}(\eta_s \phi_-^2) +e_{n-}(\eta_n \phi_-^2) \\ \nonumber
   &&+f_+\left((L_x^2 - L_y^2) \phi_+\right) + g_+\left((L_x^2+L_y^2)\phi_+^2\right)+g_-\left((L_x^2 +L_y^2) \phi_-^2\right)\\ \nonumber
   &&+h\left((L_x^2-L_y^2)\eta_s\right).
\end{eqnarray}
\end{widetext}
Here, we denote the coupling constants by lowercase Latin letters.
In the high-temperature phase of Ca$_{1-x}$La$_{x}$FeAs$_{2}$,
the space group is not tetragonal but monoclinic ($P2_1$, \#4)
and so there are various nonzero strain components
with respect to a tetragonal reference structure.
The number of symmetry operations is also reduced greatly,
and the only symmetry operation apart from identity
and translations is the screw rotation around the [010] axis.
However, this operation has a crucial
effect for the nematic transition in this compound:
It connects the two Fe atoms in the unit cell to each other
and flips the $x$ and $y$ axes.
As a result, the presence of this symmetry operation at the high
temperature phase ensures that the ferro-orbital polarization $\phi_+$ is zero.
However, there is no symmetry operation that ensures that $\phi_-$ is not zero,
and as a result, one needs to take into account an additional $\sim \phi_-$ term
in the free energy in order to consider
the lower than tetragonal symmetry of Ca$_{1-x}$La$_{x}$FeAs$_{2}$ compound.

In summary, we have listed the order parameters
that are relevant for the nematic transition in iron pnictide superconductors,
and showed that the free energy expression does not include
any terms linearly coupled to
$\phi_+$ in the compound under study, Ca$_{1-x}$La$_{x}$FeAs$_{2}$.
Therefore there will be a sharp phase transition between a phase with $\phi_+ = 0$
at high $T$ and a low $T$ phase with $\phi_+ \neq 0$
like in all other iron pnictide materials.

\section{Summary and conclusions}

We have checked the phase stability of CaFeAs$_{2}$
and Ca$_{1-x}$La$_{x}$FeAs$_{2}$ compounds.
The spacer, zig-zag As chain, nearly has the $1-$ valence state,
so that it does not form the As square-net,
which is reminiscent of the Peierls type instability.
According to the convex-hull construction,
CaFeAs$_{2}$ is above the hull with 13 meV/atom.
Further stabilization is possible with rare earth doping
in CaFeAs$_{2}$ material.
We have also calculated the optical conductivity
of Ca$_{1-x}$La$_{x}$FeAs$_{2}$ based on the DFT + eDMFT method,
and found a large in-plane resistivity anisotropy.
This large anisotropy does not originate from electronic nematicity,
but from the structural anisotropy
arising from the zig-zag As chain.
The electronic correlations do not induce but nevertheless enhance this
anisotropy, as seen from the difference of DFT and eDMFT results.
For Ca$_{0.7}$La$_{0.3}$FeAs$_{2}$ compound,
we found a low-frequency peak of 0.34 eV
in the in-plane $yy$ component of the optical conductivity.
This peak corresponds to the coherent interband transition
within the zig-zag As chain.

\begin{figure}[t]
\includegraphics[width=8.5 cm]{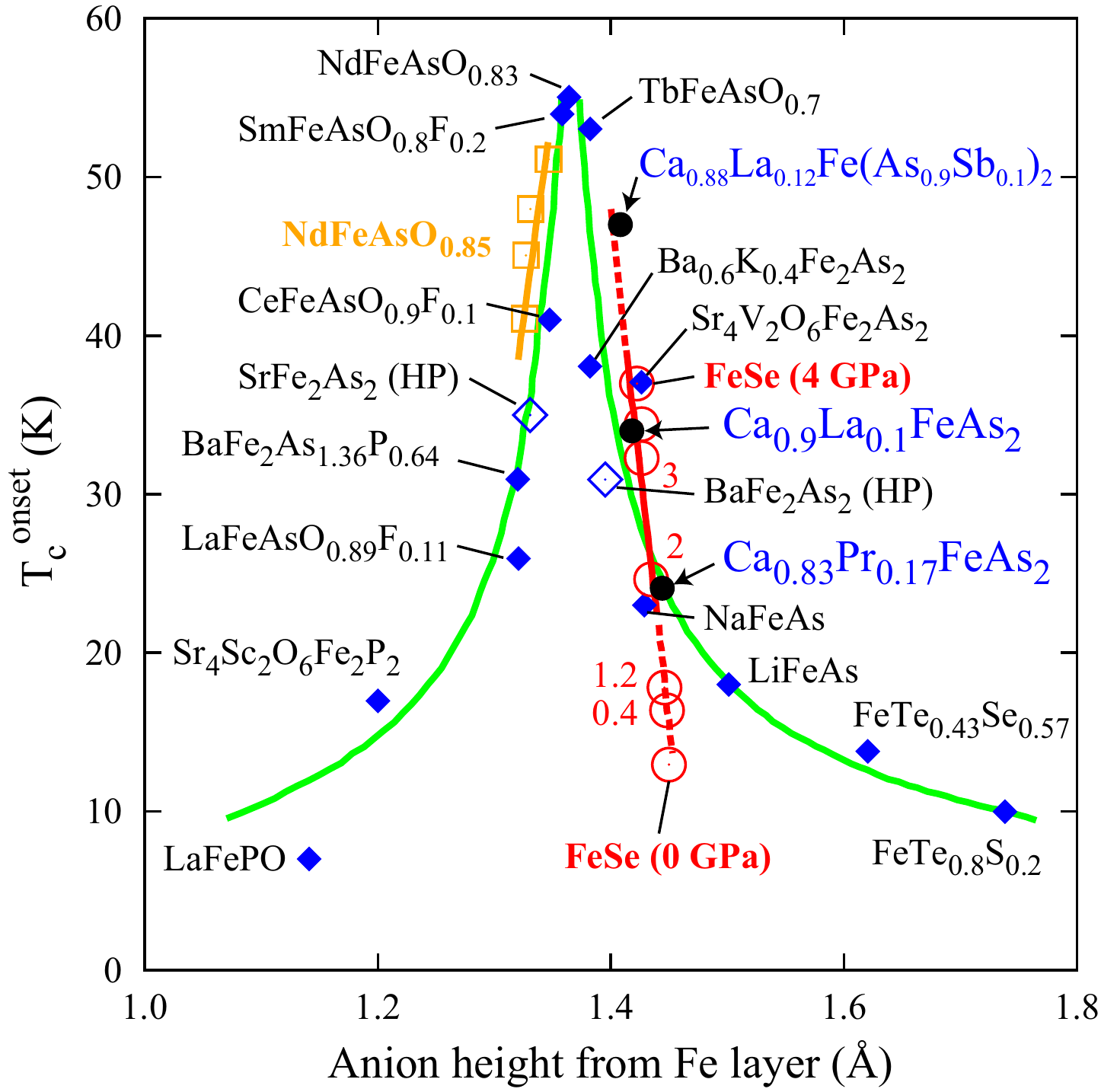}
\caption{(Color Online)
T$_c$ vs. Anion height in iron pnictide and chalcogenide superconductors.
(Reproduced with permission from Ref. \cite{Mizuguchi10}.)
The three data points we added (black) for the 112 superconductors
display a similar trend as the other iron based high T$_C$ superconductors.
This suggests that the metallic spacer layer in the 112 family
does not influence superconductivity.
}
\label{fig:Tc}
\end{figure}

In Fig. \ref{fig:Tc} we reproduce the plot of
the distance between the anion and Fe layers and T$_c$ in
iron pnictide superconductors. The three data points we add from published data
falls perfectly on the lines that were drawn to underline the striking
correlation with this single structural parameter and the superconducting
critical temperature. This suggests that despite being noncentrosymmetric and
having a metallic spacer layer, the superconductivity in the rare-earth doped
CaFeAs$_2$ is no different that in other iron pnictide superconductors. This
supports the spin-fluctuation based theories as opposed to theories that
rely on charge fluctuations.

\section{Acknowledgments}

C.-J. Kang and G. Kotliar were supported by the DOE EFRC CES program.
T. Birol acknowledges the support of the Center for Materials Theory at Rutgers University.



\begin{thebibliography}{99}
\bibitem{cal}
    We used Ising-type Coulomb interaction in the DFT + eDMFT calculations,
    as done in the early eDMFT calculation in Ref. \cite{Jiang16}.
\end{thebibliography}
\end{document}